\documentclass[reprint, superscriptaddress, amsmath,amssymb,aps, floatfix]{revtex4-2}

\usepackage{float}
\usepackage{dsfont}
\usepackage{graphicx}
\usepackage{dcolumn}
\usepackage{bm}
\usepackage{bbm}
\usepackage[utf8]{inputenc}
\usepackage{amssymb}
\usepackage{amsmath}
\usepackage{graphicx}
\usepackage{amsthm}
\usepackage{cancel}
\usepackage{mathtools}
\usepackage{upgreek}
\usepackage{xcolor}
\usepackage{placeins}
\usepackage[offset=1.3em]{simpler-wick}
\usepackage{stackengine}
\setstackEOL{\#}
\setstackgap{L}{12pt}

\usepackage{simplewick}
\usepackage{slashed}
\usepackage{hyperref}
\hypersetup{
    hidelinks,
    colorlinks=true,
    citecolor=blue,
    linkcolor=blue,
    urlcolor=blue,
    }

\allowdisplaybreaks

\newcommand{\eq}[1]{Eq.\ (\ref{#1})}
\newcommand{\fig}[2]{Fig.\ \ref{#1}#2}
\newcommand{\figpanels}[3]{Figs.\ \ref{#1}#2-\ref{#1}#3}

\newcommand{\set}[1]{\left\{ #1 \right\}}
 
\newcommand{\bra}[1]{\left\langle #1 \right|}

\newcommand{\ket}[1]{\left| #1 \right\rangle}

\newcommand{\braket}[1]{\left\langle #1 \right\rangle}
\newcommand{\braketsm}[1]{\langle #1 \rangle}

\newcommand{\mysum}{\mathrm{sum}}
\newcommand{\sumket}{\left| \mysum \right\rangle}

\usepackage[linesnumbered,ruled,vlined]{algorithm2e}
\newcommand{\mytrans}{\mathrm{trans}}
\newcommand{\myinit}{\mathrm{init}}
\newcommand{\calR}{\mathcal{R}}
\newcommand{\calM}{\mathcal{M}}
\newcommand{\calT}{\mathcal{T}}
\newcommand{\calO}{\mathcal{O}}
\newcommand{\calC}{\mathcal{C}}
\newcommand{\bbR}{\mathbb{R}}
\newcommand{\bbO}{\mathbb{O}}

\SetKwProg{Fn}{Function}{:}{}
\SetKwFunction{FInitialize}{Initialize}
\SetKwFunction{FIsRuleEligible}{IsRuleEligible}
\SetKwFunction{FFlipModeExcitation}{FlipModeExcitation}
\SetKwFunction{FGillespieStep}{GillespieStep}
\SetKwFunction{FChoose}{Choose}
\SetKwFunction{FSum}{Sum}
\SetKwFunction{FProd}{Prod}
\SetKwFunction{FExponential}{SampleExponentialDist}

\begin{document}

\preprint{APS/123-QED}

\title{Algebraic and diagrammatic methods for the rule-based modeling \\ of multi-particle complexes}

\author{Rebecca J. Rousseau}
 \email{rroussea@caltech.edu}
\affiliation{Department of Physics, California Institute of Technology, Pasadena, CA 91125}

\author{Justin B. Kinney}
 \email{jkinney@cshl.edu}
\affiliation{Simons Center for Quantitative Biology, Cold Spring Harbor Laboratory,  Cold Spring Harbor, NY 11724}

\date{\today}

\begin{abstract}
The formation, dissolution, and dynamics of multi-particle complexes is of fundamental interest in the study of stochastic chemical systems. In 1976, Masao Doi introduced a Fock space formalism for modeling classical particles. Doi's formalism, however, does not support the assembly of multiple particles into complexes. Starting in the 2000's, multiple groups developed rule-based methods for computationally simulating biochemical systems involving large macromolecular complexes. However, these methods are based on graph-rewriting rules and/or process algebras that are mathematically disconnected from the statistical physics methods generally used to analyze equilibrium and nonequilibrium systems. Here we bridge these two approaches by introducing an operator algebra for the rule-based modeling of multi-particle complexes. Our formalism is based on a Fock space that supports not only the creation and annihilation of classical particles, but also the assembly of multiple particles into complexes, as well as the disassembly of complexes into their components. Rules are specified by algebraic operators that act on particles through a manifestation of Wick's theorem. We further describe diagrammatic methods that facilitate rule specification and analytic calculations. We demonstrate our formalism on systems in and out of thermal equilibrium, and for nonequilibrium systems we present a stochastic simulation algorithm based on our formalism. The results provide a unified approach to the mathematical and computational study of stochastic chemical systems in which multi-particle complexes play an important role. 
\end{abstract}

\maketitle

\section{\label{sec:intro}Introduction}
Large complexes of classically behaving particles play a central role in a variety of scientific disciplines. For example, many essential biological processes depend on large complexes formed by proteins, nucleic acids, and/or other macromolecules. A common theme in such systems is ``combinatorial complexity'' \cite{HlavacekRules:2006}, i.e., that an immense (and often infinite) variety of molecular complexes can form from a relatively small number of interaction rules governing the assembly of a relatively small number of molecular components. Nevertheless, mathematical methods for analyzing stochastic chemical systems that exhibit such combinatorial complexity have yet to be developed.

In 1976, Masao Doi \cite{Doi:1976a,Doi:1976b} introduced a Fock space formalism for modeling many-body systems of classical particles. This approach was further developed by others \cite{Grassberger:1980, Goldenfeld:1984, Peliti:1985kp}, and has proven useful in the study of diffusion-limited aggregation \cite{Cardy:1995, Wijland:2001,MattisGlasser:1998} and other problems in statistical physics \cite{WalczakWiggins:2006,Mjolsness:2013,WeberFrey:2017}. As in quantum field theory, Doi's formalism supports the creation and annihilation of particles, but does not support the assembly of preexisting particles into complexes. Consequently, analyzing systems that involve multi-particle complexes using this formalism requires specifying one distinct field for every distinct species of complex. This makes Doi's formalism unwieldy for analyzing systems that exhibit substantial combinatorial complexity. 

\begin{figure*}[t]
    \centering
    \fbox{\includegraphics[width=0.97\textwidth]{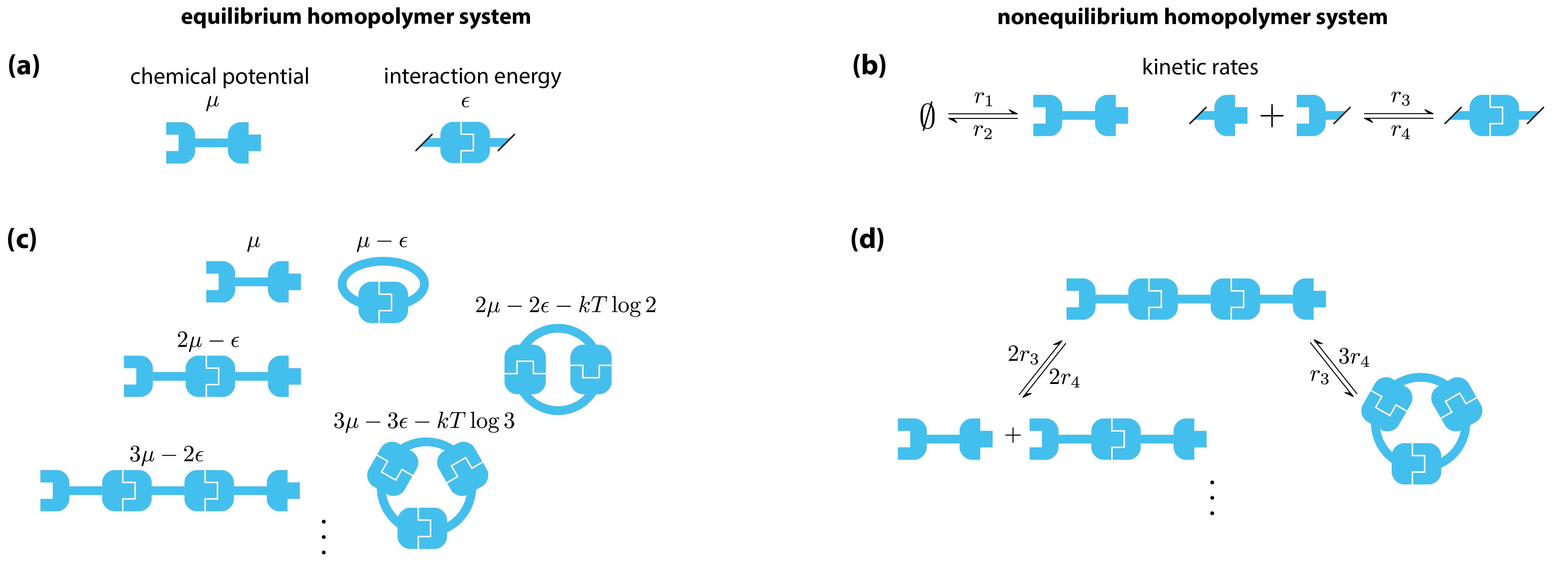}}
    \caption{Homopolymer in zero dimensions. The system comprises a single species of component particle, with each particle having two domains capable of forming heterotypic interactions. (a) In thermal equilibrium, system behavior is governed by the chemical potential ($\mu$) and interaction energy ($\epsilon$). (b) Out of equilibrium, the system is governed by the kinetic rates for monomer creation ($r_1$), monomer annihilation ($r_2$), interaction formation ($r_3$), and interaction dissolution ($r_4$). (c) The rules in panel (a) generate an infinite number of possible complexes: $x$-chains and $x$-rings for all $x=1,2,\ldots$, each complex with a distinct chemical potential. The log terms in the chemical potentials of the $x$-rings result from their rotational symmetry. (d) The rules in panel (b) result in an infinite web of reactions between different polymeric complexes, with reaction rates proportional to $r_3$ and $r_4$ in ways that depend nontrivially on the identities of the specific reactants and products. }
    \label{fig:homopolymer_cartoon}
\end{figure*}

Consider, for example, the homopolymer system illustrated in \fig{fig:homopolymer_cartoon}{}. This system comprises one type of component particle having two sites capable of forming an interaction.  In thermal equilibrium the system's behavior is governed by two quantities: the chemical potential of the particles and the energy of interaction [\fig{fig:homopolymer_cartoon}{(a)}]. Out of equilibrium the system is governed by four rate parameters describing the appearance and disappearance of particles, as well as their mutual binding and unbinding  [\fig{fig:homopolymer_cartoon}{(b)}]. But despite how simple this system is to describe in words and pictures, modeling this system in Doi's formalism is complicated because the above rules lead to an infinite number of possible polymeric complexes. To apply Doi's formalism, one must define an infinite number of fields, one for every species of complex. For equilibrium systems, one must then manually specify the chemical potential of each species [\fig{fig:homopolymer_cartoon}{(c)}]. For systems out of equilibrium, one must manually specify the rate of reaction between all reacting sets of species [\fig{fig:homopolymer_cartoon}{(d)}]. And in doing so, one must take care that the chemical potentials and/or reaction rates written down are expressed correctly as functions of the underlying model parameters, as Doi's formalism provides no means of computing these quantities.

Clearly something is missing. Ideally, the formalism one uses to describe systems of multi-particle complexes should allow one to mathematically derive the set of possible complexes, the chemical potentials of each complex, and the rates of  reaction between complexes from the underlying rules and their associated parameters. This paper develops a formalism that does this.

The problem of combinatorial complexity has long been recognized in the field of computational systems biology. Starting in the 2000s, researchers studying biological signaling pathways began developing ``rule-based'' approaches for simulating chemical systems of  multi-particle complexes \cite{Regev:2001, BlinovBNG:2004, Baeten:2005, HlavacekRules:2006, Faeder:2009BNG, FeretKappa:2009, SneddonNFSim:2011, ChylekRules:2014}. Some of these efforts have produced sophisticated software ecosystems, such as BioNetGen \cite{BlinovBNG:2004, Faeder:2009BNG, SneddonNFSim:2011,Harris:2016} and Kappa \cite{FeretKappa:2009,BoutillierKappa:2018}. These simulation approaches, however, are based on formal representations that do not lend themselves to analytical calculations using the mathematical methods of statistical physics. For example, BioNetGen is based on a process algebra describing the formation and dynamics of port graphs (i.e., graphs with edges attached through ports) \cite{AndreiPort:2008} while Kappa is based on the $\upkappa$-calculus process algebra \cite{FeretKappa:2009,BoutillierKappa:2018}. As a result, work in this area has been confined to computational analyses rather than analytic calculations. 

Here we bridge the divide between Doi's mathematical formalism and rule-based methods for computationally modeling biochemical systems. Using a Fock space for classical particles reminiscent of but distinct from that of Doi, we develop an operator algebra that allows not only for the creation and annihilation of particles, but also for the assembly of particles into complexes. We show that this operator algebra allows one to mathematically analyze equilibrium and nonequilibrium systems that are defined in a rule-based manner, and can also be used as a basis for computational analyses using stochastic simulations. 

After introducing how microstates and macrostates are represented, we apply the formalism to three systems in thermal equilibrium: a monomer system, a homodimer system, and a homopolymer system. Next we show how our formalism can be used to define and analyze nonequilibrium systems, both through the analytic derivation of master equations and through computational analysis carried out using a stochastic simulation algorithm. We end by illustrating the versatility of our formalism, showcasing the variety and complexity of system behavior that can arise from positing different sets of rules. 

\section{\label{sec:foundations}Foundations}
\subsection{Microstates}

Following Doi we define a set $\mathcal{S}$ of microstates where each microstate $s \in \mathcal{S}$ corresponds to a unit vector $\ket{s}$. The resulting set of pure states forms an orthonormal basis for the Fock space. The vector $\ket{\psi}$ describing the system is then given by a probabilistic mixture of pure states, 
\begin{equation}
    |\psi\rangle = \sum_{s\in\mathcal{S}}p_{s}|s\rangle,\label{eq:psi1}
\end{equation}
where $p_s$ represents the probability of the system being in state $s$. It is useful to define the sum of all possible states as the ``sum vector''
\begin{equation}
    \sumket = \sum_{s\in\mathcal{S}}\ket{s}.
\end{equation}
The expectation value of any operator $\mathbb{O}$ is then given by $\braket{\text{sum}|\mathbb{O}|\psi}$, and probability normalization requires that $  \braket{\textrm{sum} | \psi} = 1$.

The Fock space supports systems both in and out of thermal equilibrium. In equilibrium, the state vector $\ket{\psi}$ for the system (which is taken to be in the grand canonical ensemble) can be expressed in terms of a Hamiltonian operator $\mathbb{H}$ that assigns a free energy to each microstate:
\begin{equation}
    |\psi\rangle = \frac{e^{-\beta\mathbb{H}}}{Z}|\text{sum}\rangle,~~~\textrm{where}~~~
    Z = \langle\text{sum}|e^{-\beta\mathbb{H}}|\text{sum}\rangle  \label{eq:boltzmann_equation}
\end{equation}
is the partition function, $\beta = 1/k_{B}T$ where $k_B$ is Boltzmann's constant and $T$ is temperature, and $\mathbb{H} \ket{s} = H_s \ket{s}$ where $H_s$ denotes the free energy of state $s$. The dynamics of the system state $\ket{\psi}$ out of equilibrium is described by 
\begin{equation}
    \frac{d}{dt}\ket{\psi} = \mathbb{W} \ket{\psi}, \label{eq:microstate_master_equation}
\end{equation}
where $\mathbb{W}$ is a transition operator. We call this the ``macrostate master equation.'' In terms of the scalar transition rates $W_{s \to t}$ from microstate $s$ to microstate $t$, the transition operator is 
\begin{equation}
     \mathbb{W} = \sum_{s,t \in \mathcal{S}} W_{s\to t} \left(  \ket{t}\!\bra{s} - \ket{s}\!\bra{s} \right). \label{eq:transition_operator}
\end{equation}
Note that the first term in the summand reflects the flow of probability into $p_t$, while the second term reflects the flow of probability out of $p_s$. We call these the ``reaction'' and ``depletion'' terms, respectively.

\subsection{Macrostates}

Unlike in Doi's formalism, the microstates in our formalism represent not only the externally observable properties of particles, but also their unobservable internal states. These internal states are, in fact, what make particles in our formalism identifiable and thus allow complexes to be constructed from preexisting particles. We therefore distinguish between the microstates of a system (represented by the $\ket{s}$ vectors) and the macrostates of the system. 

In this work we focus on zero-dimensional (i.e., well-mixed) populations of particles and complexes. Assuming there are $K$ possible observably distinct species of complex, each macrostate is characterized by a vector $\vec{n} = (n_1, \ldots, n_K)$ where each $n_k$ quantifies the number of complexes of species $k$. The corresponding macrostate vector is defined to be the sum of all microstate vectors consistent with the macrostate, i.e.,
\begin{equation}
    \ket{\vec{n}} = \sum_{ s | \vec{n}}  \ket{s}.
\end{equation}
Note that $\sumket = \sum_{\vec{n}} \ket{\vec{n}}$, and that the probability of a the system being in a macrostate $\vec{n}$ given $\ket{\psi}$ is
\begin{equation}
P(\vec{n}) = \braketsm{\vec{n} | \psi}. \label{eq:Pn_from_inner_product}
\end{equation}
In equilibrium systems, each macrostate is an eigenstate of the Hamiltonian:
\begin{equation}
	\mathbb{H} \ket{\vec{n}} = - \sum_k n_k \mu_k \ket{\vec{n}},
\end{equation}
where $\mu_k$ is the (bare) chemical potential of species $k$. Consequently, the probability of the system having macrostate $\vec{n}$ is a product of species-specific Poisson distributions, i.e.,
\begin{equation}
	P(n_k) = \frac{1}{Z_k} \frac{e^{\beta \mu_k n_k}}{n_k!},~~\textrm{where}~~Z_k = \exp \left[ e^{\beta \mu_k} \right].
\end{equation}
In nonequilibrium systems, this probability becomes a function of time $t$ and evolves according to the ``macrostate master equation''
\begin{equation}
    \frac{d}{dt} P_t(\vec{n}) = \sum_{\vec{n}'} W_{\vec{n}\vec{n}'} P_t(\vec{n}'),
\label{eq:macrostate_master_equation}
\end{equation}
where $W_{\vec{n}\vec{n}'} = \bra{\vec{n}} \mathbb{W} \ket{\vec{n}'} / \braket{\vec{n}'|\vec{n}'}$ are the macrostate-specific transition rates. In later sections we compute these $W_{\vec{n}\vec{n}'}$ from the transition operator $\mathbb{W}$, but  instead of calculating each rate directly we find it simpler to calculate the vector
\begin{align}
    \ket{J(\vec{n})} = \mathbb{W}^\dagger \ket{\vec{n}}. \label{eq:J_vec_def}
\end{align}
We call $\ket{J(\vec{n})}$ the ``flux projector'' since taking the inner product of it with $\ket{\psi}$ yields a vector of probability fluxes, i.e., $\dot{P}_t(\vec{n}) = \braket{J(\vec{n}) | \psi(t)}$.

\section{\label{sec:monomer}Monomer in equilibrium}

\subsection{Microstates and macrostates}

We now construct the Fock space on which our formalism is based, using a system of monomeric particles for concreteness. The particles are represented using a hard-core boson field, $A$, which is assumed to have $N$ excitation modes. Each mode $A_i$ is indexed by a number $i \in \mathcal{N} = \set{1, \ldots, N}$ that represents the internal state of a particle. This index allows the formalism to track individual particles that are outwardly identical. In what follows we keep $N$ finite for concreteness, but all physically meaningful calculations are performed in the $N \to \infty$ limit. 

Each mode $A_i$ can be in one of two orthonormal states: $\ket{1}_i$ represents the presence of a particle with internal state $i$, while $\ket{0}_i$ represents its absence. Microstates are given by tensor products over all modes. Specifically, a microstate representing $K$ particles having indices $\mathcal{I} \subseteq \mathcal{N}$ is represented by
\begin{equation}
    \ket{\mathcal{I}} = \bigotimes_{i\in\mathcal{N}} \left\{ 
        \begin{array}{cl} 
            \ket{1}_i & \textrm{if}~i \in \mathcal{I}, \\ 
            \ket{0}_i & \textrm{otherwise}.
        \end{array}
    \right. \label{eq:ket_I_def}
\end{equation}
These states are orthonormal, i.e., $\braket{\mathcal{I} | \mathcal{J}} 
    = \delta_{\mathcal{I} \mathcal{J}}$. The resulting macrostates of the system are 
\begin{equation}
\ket{n} = \sum_{\mathcal{I} : |\mathcal{I}| = n}\ket{\mathcal{I}},~~~\textrm{for}~~~n=0,\ldots,N.
\end{equation}
The vacuum state, $\ket{0} = \ket{\emptyset}$, is both a microstate and a macrostate.

These definitions readily extend to systems defined by multiple fields. Consider  a mixture of monomer species $A$ and $B$. The microstate comprising $A$ monomers having indices $\mathcal{I}$ and $B$ monomers having indices $\mathcal{J}$ is given by
\begin{eqnarray}
    \ket{\mathcal{I}, \mathcal{J}}_{A,B} 
         = 
        \ket{\mathcal{I}}_A 
        \otimes
        \ket{\mathcal{J}}_B , 
\end{eqnarray}
where the subscripts indicate the Fock space in which each state vector lives. The corresponding macrostates and sum states are given by analogous tensor products. Systems with three or more fields are defined similarly.

\subsection{Mode and field operators}\label{subsec:modefieldops}
We now define four types of mode-specific operators: creation, annihilation, presence, and absence. The creation operator for mode $i$ is defined to be $\hat{A}_i = \ket{1}_i\bra{0}_i$. When applied to a microstate $\ket{\mathcal{I}}$, this operator has the effect
\begin{equation}
    \hat{A}_i \ket{\mathcal{I}} = \left\{ 
        \begin{array}{cl} 
            0 & \textrm{if}~i \in \mathcal{I}, \\
            \ket{\mathcal{I} \cup \set{i}} & \textrm{otherwise}.
        \end{array}
    \right.
\end{equation}
The corresponding annihilation operator is defined to be $\check{A}_i = \hat{A}_i^\dagger$, and has the effect
\begin{equation}
    \check{A}_i \ket{\mathcal{I}} = \left\{ 
        \begin{array}{cl} 
            \ket{\mathcal{I} \setminus \set{i}} & \textrm{if}~i \in \mathcal{I}, \\
            0 & \textrm{otherwise}.
        \end{array}
    \right.
\end{equation}
The presence operator is defined as $\bar{A}_i = \hat{A}_i \check{A}_i$. $\bar{A}_i\ket{\mathcal{I}}$ is one if $i \in \mathcal{I}$ and zero otherwise. The absence operator is defined to be $\tilde{A}_i = \hat{A}_i \check{A}_i = 1 - \bar{A}_i$. Note that $\bar{A}_i$ and $\tilde{A}_i$ are self-adjoint. 

We highlight several key algebraic properties of these operators. First, creation and annihilation operators are nilpotent, i.e., $\hat{A}_i^2 = \check{A}_i^2 = 0$. Second, the commutator 
\begin{equation}
    [\check{A}_i, \hat{A}_j] = \delta_{ij}(1 - 2 \bar{A}_i)
\end{equation}
is very different than one finds in the harmonic oscillator algebra, and thus in other algebras used to model classical particles \cite{Doi:1976a, Peliti:1985kp}. Third, the commutator
\begin{equation}
    ~[\bar{A}_{i},\hat{A}_{j}] = \delta_{ij}\hat{A}_{j},\label{eq:bar_hat_commutation_relation}
\end{equation}
plays an important role later when constructing multi-particle complexes from component particles. Appendix \ref{app:operatoralgebra} lists some additional useful properties.

We further define field-specific creation, annihilation, presence, and absence operators as sums over all of the corresponding mode operators, i.e.,
\begin{align}
    \hat{A} &= \sum_i \hat{A}_i,
\end{align}
and similarly for $\check{A}$, $\bar{A}$, and $\tilde{A}$. These field operators satisfy the useful commutation relations
\begin{equation}
    [\check{A}, \hat{A}] = N - 2 \bar{A},~~~~~~~~~~~[\bar{A},\hat{A}] = \hat{A}.
\end{equation}
Macrostates are given by
\begin{equation}
   \ket{n} = \frac{\hat{A}^n}{n!} \ket{0}. \label{eq:ket_n}
\end{equation}
Here the combinatorial factor corrects for each set of modes being summed over $n!$ times in the operator product $\hat{A}^n$. Note that $\ket{n} = 0$ if $n > N$, since this would cause every term in $\hat{A}^n$ to contain at least one factor of $\hat{A}_i^2$. Applied to a macrostate, one finds that
\begin{align}
	\hat{A} \ket{n} &= (n\!+\!1) \ket{n\!+\!1},~~~~~~~~~\bar{A} \ket{n} = n \ket{n}, \label {eq:Ahatn_Abarn} \\
	\check{A} \ket{n} &= (N\!-\!n\!+\!1) \ket{n\!-\!1},~~~\tilde{A} \ket{n} = (N\!-\!n) \ket{n}. \label{eq:Acheckn_Atilden}
\end{align}
See Appendix \ref{app:operatoralgebra} for a derivation of these results.

In the large $N$ limit, the number of modes that are excited in any macrostate $\ket{n}$ with substantial physical probability becomes negligible compared to $N$. In what follows we therefore approximate \eq{eq:Acheckn_Atilden} as 
\begin{align}
	\check{A} \ket{n} \approx N \ket{n-1},~~~~~\tilde{A} \ket{n} \approx N \ket{n}.
\end{align}
By similar logic we can also approximate $[\check{A}, \hat{A}] \approx N$, etc.

 Finally, it is useful to consider the coherent state,
\begin{equation}
\ket{z} = \sum_{n=0}^\infty z^n \ket{n} = e^{zA}\ket{0}.
\end{equation}
This allows one to express the generating function for the distribution over macrostates as $\braket{z|\psi}$. Note that setting $z=1$ recovers the sum state, i.e., $\ket{1} = \sumket$.

\subsection{Hamiltonian operator}\label{sec:monomerHam}
In what follows we assume that each system of interest is contained within a volume $V$. For a gas of monomers, the relevant Hamiltonian is $\mathbb{H} = -\mu\bar{A}$, where $\mu$ denotes a bare chemical potential. We use the term ``bare'' to emphasize that $\mu$ determines the excitation probability for each independent mode $A_i$, whereas the concentration of monomers depends on $\mu$, $N$, and $V$. The generating function for the equilibrium probability distribution over macrostates is found by:
\begin{align}
    \braket{z | \psi} &= \frac{1}{Z} \bra{0} e^{z\check{A}} e^{\beta \mu \bar{A}} e^{\hat{A}} \ket{0} \\
    &= \frac{1}{Z} \bra{0} e^{z\check{A}} e^{\lambda \hat{A}} \ket{0}~~(\textrm{defining}~~\lambda = e^{\beta \mu}) \\
    &= \frac{1}{Z} \prod_i \bra{0}_i e^{z\check{A}_i} e^{\lambda \hat{A}_i} \ket{0}_i \\
    &= \frac{1}{Z} \prod_i \bra{0}_i (1 + z \check{A}_i) (1 + \lambda \hat{A}_i) \ket{0}_i \\
    &= \frac{1}{Z} \left( 1 + z \lambda \right)^N \\
    &= \left( \frac{1 + z \lambda}{1 + \lambda} \right)^N. 
\end{align}
In the first step we used the fact that $f(\bar{A})g(\hat{A})\ket{0} = g(f(1)\hat{A})\ket{0}$ for any functions $f$ and $g$. The resulting quantity $\lambda$ is the per-mode fugacity corresponding to chemical potential $\mu$. In the second step we used the fact that operators for different modes commute. In the third step we used the nilpotency of $\hat{A}_i$ to truncate the expansions of each exponential. Finally, we used the normalization requirement $\braket{1 | \psi} = 1$ to determine the partition function $Z = (1+\lambda)^N$. The result is the generating function for the binomial distribution corresponding to $N$ modes with a per-mode excitation probability of $\lambda / (1 + \lambda)$. From this generating function we find that the expected concentration of monomers is  
\begin{equation}
    \frac{\braket{\bar{A}}}{V} = \frac{1}{V}\left. \frac{d}{dz}  \braket{z | \psi} \right|_{z=1} \!\!\!\!=~ \frac{1}{V}\frac{N \lambda}{1 +\lambda} = \frac{\lambda'}{1 +V\lambda'/N},
\end{equation}
where $\lambda' = \frac{N}{V}\lambda$. Keeping $\braket{\bar{A}}/V$ constant while taking $N \to \infty$ requires holding $\lambda'$ approximately constant and thus rescaling $\lambda \sim \frac{V}{N}$. In this limit we get
\begin{align}
    \braket{z | \psi} = e^{(z-1) V \lambda'}~~\Rightarrow~~\ket{\psi} = e^{-V \lambda'} \ket{V \lambda'}. \label{eq:monomer_solution}
\end{align}
The corresponding partition function is $Z = e^{V \lambda'}$. Note that $\braket{z | \psi}$ is the generating function for a Poisson distribution with mean $V \lambda'$. We thus see that 
\begin{align}
    \mu' = k_B T \log \lambda' = \mu + k_B T \log \frac{N}{V}
\end{align}
is the effective chemical potential, i.e., the chemical potential appropriately renormalized to account for the $N$ modes available for excitation in volume $V$. It is therefore $\mu'$, not $\mu$, that reflects the physically measurable chemical potential.

\section{\label{sec:homodim}Homodimer in equilibrium}
\subsection{Composite operators}

Multi-particle complexes are represented as products of mode operators for three kinds of fields: particle fields, interaction fields, and site fields. For example, we define the creation operator for a dimer of two $A$ particles by the composite operator
\begin{equation}
    \hat{D}_{ij} = \hat{I}_{ij}\hat{a}_{i}\hat{a}_{j}\hat{A}_{i}\hat{A}_{j}, \label{eq:dimer_def}
\end{equation}
which is the product of mode operators for a particle field $A$, an interaction field $I$, and a site field $a$. More specifically, $\hat{A}_{i}$ and $\hat{A}_{j}$ create the two component particles, $\hat{I}_{ij}$ registers that these two particles interact with one another, and $\hat{a}_i$ and $\hat{a}_j$ respectively indicate that the $A_i$ and $A_j$ particles are each participating in an interaction and are therefore not free to interact with additional particles. Note that the index of the dimer creation operator is the pair of monomer indices, $(i,j)$. Since the monomer is symmetric, we assume that $\hat{I}_{ij} = \hat{I}_{ji}$, and so $\hat{D}_{ij} = \hat{D}_{ji}$. Note also that $\hat{D}_{ii} = 0$ because of the nilpotency of $\hat{A}_i$ and $\hat{a}_i$. The number of internal states for the dimer is therefore $N_D = {N \choose 2} \approx N^2/2$. The dimer annihilation, presence, and absence operators are defined in terms of the creation operator in the same manner as for a single particle:
\begin{align}
	\check{D}_{ij} &= \hat{D}_{ij}^\dagger = \check{I}_{ij}\check{a}_{i}\check{a}_{j}\check{A}_{i}\check{A}_{j}, \\
	\bar{D}_{ij} &= \hat{D}_{ij}\check{D}_{ij} = \bar{I}_{ij}\bar{a}_{i}\bar{a}_{j}\bar{A}_{i}\bar{A}_{j}, \\
	\tilde{D}_{ij} &= \check{D}_{ij}\hat{D}_{ij} = \tilde{I}_{ij}\tilde{a}_{i}\tilde{a}_{j}\tilde{A}_{i}\tilde{A}_{j}. 
\end{align}
The field operator $\hat{D}$ is given by $\hat{D} = \frac{1}{2} \sum_{i,j} \hat{D}_{ij}$, where the factor $1/2$ compensates for double counting in the sum. The field operators $\check{D}$, $\bar{D}$, and $\tilde{D}$ are defined similarly. 

The homodimer system also comprises free monomers. We represent these by a separate composite field $M$ defined by the mode operator $\hat{M}_i = \hat{A}_i \tilde{a}_i$. The corresponding number of internal states is $N_M = N$, and the three related mode operators are $\check{M}_i = \check{A}_i \tilde{a}_i$, $\bar{M}_i = \bar{A}_i \tilde{a}_i$, and $\tilde{M}_i = \tilde{A}_i \tilde{a}_i$. The corresponding field operators are defined as sums over $i$. Note the inclusion of $\tilde{a}_i$ in $\hat{M}_i$ ensures that $\bar{M}$ does not count $A$ particles that are components of dimers, $\check{M}$ does not annihilate such particles (which would leave dangling $I$ and $a$ modes), etc. 

The macrostate comprising $m$ monomers and $d$ dimers is given by
\begin{equation}
	\ket{m, d} = \frac{\hat{M}^{m}}{m!} \frac{\hat{D}^{d}}{d!} \ket{0}. \label{eq:gallery_representation_of_expansion}
\end{equation}
Using $[\tilde{a}_i, \hat{a}_j] = -\delta_{ij} \hat{a}_i$ and $\hat{A}_i^2 = 0$, one can readily verify that $\hat{M}$ and $\hat{D}$ commute. The corresponding coherent state is therefore 
\begin{equation}
	\ket{z_M, z_D} = \sum_{m=0}^{\infty} \sum_{d=0}^{\infty} z_M^m z_D^d \ket{m, d} = e^{z_M \hat{M} + z_D \hat{D}} \ket{0}.\label{eq:coherent_dim}
\end{equation}

\subsection{Sectoring by species} \label{subsec:sectoring}

Now consider a Hamiltonian in which each $A$ particle has chemical potential $\mu$ and each interaction has Gibbs free energy $\epsilon$:
\begin{equation}
    \mathbb{H} = -\mu \sum_{i}\bar{A}_{i}+\epsilon\frac{1}{2}\sum_{i,j}\bar{I}_{ij}.\label{eq:homodimH}
\end{equation}
To compute the equilibrium state of the system, we re-express the Hamiltonian as a sum of terms that operate separately on monomers and dimers. Using the identity $1 = \tilde{a}_i + \bar{a}_i$, we split the Hamiltonian into two parts, $\mathbb{H} = \mathbb{H}_M + \mathbb{H}_D$, where
\begin{align}
    \mathbb{H}_M\!=\!-\mu\!\sum_i\!\bar{A}_i \tilde{a}_i,~~\mathbb{H}_D\!=\!-\mu \sum_i\!\bar{A}_i \bar{a}_i\!+\!\frac{\epsilon}{2} \sum_{i,j} \bar{I}_{ij}.
\end{align}
These operators satisfy the commutation relations 
\begin{align}
 &[\mathbb{H}_M, \hat{M}] = -\mu_M \hat{M},
 ~~~~
 [\mathbb{H}_D, \hat{D}] = -\mu_D\hat{D}, \nonumber \\
&[\mathbb{H}_M, \hat{D}] = 0, 
~~~~~~~~~~~~~
[\mathbb{H}_D, \hat{M}] = 0, \label{eq:dimer_H_comms}
\end{align}
where  $\mu_M = \mu$ and $\mu_D = 2 \mu - \epsilon$ are the bare monomer and dimer chemical potentials. Next we compute the generating function:
\begin{align}
    \braket{z_M, z_D | \psi} &= Z^{-1} \bra{0} e^{z_M \check{M} + z_D \check{D}} e^{-\beta (\mathbb{H}_M+\mathbb{H}_D)} e^{\hat{M} + \hat{D}} \ket{0} \nonumber \\
    &= Z^{-1} \bra{0} e^{z_M \check{M} + z_D \check{D}} e^{\lambda_M \hat{M} + \lambda_D \hat{D}} \ket{0} \nonumber \\
    &\approx Z^{-1} \bra{0} e^{z_M \check{M}} e^{\lambda_M \hat{M}} e^{z_D \check{D}} e^{\lambda_D \hat{D}} \ket{0}. \label{eq:dimer_generating_function_pt1}
\end{align}
In the first step we used \eq{eq:dimer_H_comms} and defined the fugacities $\lambda_M = e^{\beta \mu_M}$ and $\lambda_D = e^{\beta \mu_D}$. The second step follows from the approximation (see Appendix \ref{app:noncommutation})
\begin{equation}
    [\hat{M},\check{D}] = \frac{1}{2} \sum_{i,j} (\hat{M}_i + \hat{M}_j) \check{D}_{ij}  \approx 0. \label{eq:Mhat_Dcheck_commutator}
\end{equation}
This commutator is not exactly zero because annihilating a dimer frees up $A$ modes that can be used to create two monomers. But by way of comparison,
\begin{equation}
    [\hat{M},\check{M}] = N_M - 2 \bar{M}~~\textrm{and}~~[\hat{D},\check{D}] = N_D - 2 \bar{D}
\end{equation}
have terms that scale as $N$ and $N^2$, respectively. The effect of the commutator $[\hat{M},\check{D}]$ on a physical state is consequently negligible in the large $N$ limit. This reflects the number of modes available to create a monomer not being limiting in the physically meaningful regime. 

Next we insert a copy of the identity operator, 
\begin{equation}
    \mathbbm{1} = \sum_{m,d} \frac{\ket{m,d}\bra{m,d}}{m!d!},
\end{equation}
into the right-hand side of \eq{eq:dimer_generating_function_pt1} and observe that only the $\ket{0}\bra{0}$ term survives. Consequently,
\begin{align}
    \braket{z_M, z_D | \psi}
    &\approx Z^{-1} \bra{0} e^{z_M \check{M}} e^{\lambda_M \hat{M}} \ket{0}\bra{0} e^{z_D \check{D}}e^{\lambda_D \hat{D}} \ket{0} \nonumber \\
    &=Z^{-1} \braket{z_M| \lambda_M}\braket{z_D| \lambda_D}.
\end{align}
Setting $\braket{1,1|\psi} = 1$ we find that $Z \approx Z_M Z_D$, where $Z_M$ and $Z_D$ are the respective partition functions for the monomer and dimer species. We thus obtain
\begin{equation}
	\ket{\psi} \approx \ket{\psi}_M \otimes \ket{\psi}_D, \label{eq:dimer_sectoring}
\end{equation}
where $\ket{\psi}_M$ describes a monomer-only system, $\ket{\psi}_D$ describes a dimer-only system, and both have the same Poisson form as in \eq{eq:monomer_solution}. 

There are two important caveats to the result in \eq{eq:dimer_sectoring}. First, the sectors for distinct species are only independent in the $N \to \infty$ limit. For example, the right-hand side of \eq{eq:dimer_sectoring} has nonzero $\ket{m} \otimes \ket{d}$ terms for all values of $m \leq N$ and $d \leq N/2$, whereas each $\ket{m,d}$ term on the left-hand side is nonzero only if $m + 2d \leq N$. Second, this sectoring result holds only in equilibrium systems; indeed, the populations of particles in different sectors will generally be coupled out of equilibrium. 

Finally we discuss the scaling behavior of the system with $N$ and $V$. As in the previous section, requiring the concentration of monomers $\braket{\bar{M}}/V$ to be constant as $N \to \infty$ reveals an effective monomer chemical potential of $\mu_M' = \mu_M -+k_B T \log \frac{N_M}{V}$. Similarly, requiring the concentration of homodimers $\braket{\bar{D}}/V$ to be constant as $N \to \infty$ reveals an effective monomer chemical potential of $\mu_D' = \mu_D + k_B T \log \frac{N_D}{V}$. These relations are realized by renormalizing the parameters of the Hamiltonian so that
\begin{equation}
    \mu' = \mu + k_BT \log \frac{N}{V} ~~~\textrm{and}~~~\epsilon' = \epsilon - k_B T \log V
\end{equation}
are held constant. In terms of these quantities, the effective dimer chemical potential is $\mu'_D = 2 \mu' - \epsilon' - k_B T \log 2$, where the logarithmic term accounts for the symmetry of the molecule. This system is therefore exactly renormalizable. We thus see that, to maintain a fixed concentration of dimers, the bare interaction energy $\epsilon$ must become weaker as system volume increases. This makes sense: if $V$ increases while monomer concentration stays fixed, the number of monomers available to bond to a given monomer will increase in proportion to $V$. To keep the probability of the given monomer forming a dimer constant, the bare interaction energy $\epsilon$ must weaken as $V$ increases so that $e^{-\beta \epsilon} \propto V^{-1}$. This implies that $e^{-\beta \epsilon'} = V e^{-\beta \epsilon}$ will be fixed.

\subsection{Gallery operators}

Consider more generally a system that realizes $K$ distinct species of complex. Let $\hat{G}_k$ denote the creation operator for complex $k$ and assume that
\begin{equation}
	[\bar{G}_k, \hat{G}_{k'}] = \delta_{kk'} \hat{G}_k.
\end{equation}
We refer to the vector $\vec{\mathbb{G}} = (\hat{G}_1, \ldots, \hat{G}_k)^\top$ as the ``gallery,'' as it exhibits creation operators for all possible complexes. The gallery allows us to define the coherent state
\begin{equation}
    \ket{\vec{z}} = \exp \left\{ \vec{z}^\top \vec{\mathbb{G}} \right\} \ket{0}, \label{eq:general_coherent_state}
\end{equation}
where $\vec{z} = (z_1, \ldots, z_K)^\top$ is a vector of scalars. As in the monomer and homodimer systems, the generating function for a system $\ket{\psi}$ is $\braket{\vec{z}|\psi}$, and the sum state is
\begin{equation}
    \sumket = e^{\sum_k \hat{G}_k} \ket{0} = | \vec{1} \rangle. \label{eq:gallery_sum_state}
\end{equation}
The macrostates of the system are given by
\begin{equation}
\ket{n_1, \ldots, n_K} = \left[ \prod_{k=1}^K \frac{\hat{G}_k^{n_k}}{n_k!} \right] \ket{0},
\end{equation} 
and the effects of the four field operators on macrostates are
\begin{align}
	\hat{G}_k \ket{n_1, \ldots, n_K}\!&= (n_k\!+\!1) \ket{n_1, \ldots, n_k\!+\!1, \ldots, n_K},  \nonumber \\
	\bar{G}_k \ket{n_1, \ldots, n_K}\!&= n_k \ket{n_1, \ldots, n_K}, \nonumber\\
	\check{G}_k \ket{n_1, \ldots, n_K}\!&\approx N_k \ket{n_1, \ldots, n_k\!-\!1, \ldots, n_K}, \nonumber \\
	\tilde{G}_k \ket{n_1, \ldots, n_K}\!&\approx N_k \ket{n_1, \ldots, n_K}.\label{eq:Gk_on_macrostates} 
\end{align}
where $N_k$ is the number of internal states for species $k$. 

Because different complexes can share internal components, the Hamiltonian can be defined in a rule-based manner as in \eq{eq:homodimH} instead of on a species-by-species basis. As we will see in later sections, such rule-based definitions can require far fewer than $K$ terms. If there are no energetic interactions between separate complexes, the Hamiltonian can then be equivalently expressed as 
\begin{equation}
	\mathbb{H} \simeq -\sum_{k} \mu_k \bar{G}_k. \label{eq:hamiltonian_assumption}
\end{equation}
where $\mu_k$ is the bare chemical potential for species $k$. The generating function for the equilibrium state then factorizes, i.e.
\begin{equation}
    \ket{\psi} = \bigotimes_k \ket{\psi}_k,~~\textrm{where}~~\ket{\psi}_k = e^{-V\lambda'_k} \ket{V \lambda'_k},
\end{equation}
  and where $\lambda'_k=e^{\beta \mu'_k}$ and $\mu'_k = \mu_k + k_{B}T \log \frac{N_k}{V}$ are the effective fugacity and  chemical potential of species $k$. We note that it may or may not be possible to renormalize the parameters of the Hamiltonian so that all the effective chemical potentials are independent of $V$. For example, this is possible in the homodimer system, but not in the homopolymer system discussed in the next section. 

\subsection{Factory operators}

Hamiltonians of the form in \eq{eq:hamiltonian_assumption} describe systems of non-interacting particles and might understandably be viewed as trivial. They become less trivial, however, in systems comprising large (or infinite) numbers of distinct complexes, each complex having a chemical potential that is a function of the parameters used to define the Hamiltonian. In such systems, merely enumerating different species of complex and determining their chemical potentials can be nontrivial. It is therefore natural to instead define the set of possible complexes implicitly by specifying the rules for their construction, and to use these rules to then compute the different species of complex and their associated chemical potentials. We now show how our formalism enables this.

In the case of the homodimer, the sum state can be expressed as
\begin{equation}
	\sumket = e^{\mathbb{F}_2} e^{\mathbb{F}_1} \ket{0}, \label{eq:factory_form_of_homodimer_sum_vector}
\end{equation}
where
\begin{align}
    \mathbb{F}_1 = \sum_i \hat{A}_i \tilde{a}_i,~~~~~~~
\mathbb{F}_2 = \frac{1}{2} \sum_{i,j} \hat{I}_{ij}\hat{a}_{i}\hat{a}_{j}\bar{A}_{i}\bar{A}_{j}.
\end{align}
Here, $\mathbb{F}_1 = \hat{M}$ creates free monomers, while $\mathbb{F}_2$ joins two monomers into a dimer. Specifically, $\mathbb{F}_2$ tests for the presence of two particles, $A_i$ and $A_j$, and if these already exist it joins them into a dimer $D_{ij}$. Note that neither $A$ particle can be part of an existing dimer due to the excitation of site fields $a_i$ and $a_j$. For example, 
\begin{align}
    \mathbb{F}_2 \frac{\mathbb{F}_{1}^2}{2} \ket{0} &= \frac{1}{4}\sum_{i,j,k,l}\hat{I}_{ij}\hat{a}_{i}\hat{a}_{j}\bar{A}_{i} \bar{A}_{j} \hat{A}_{k} \tilde{a}_i \hat{A}_{l} \tilde{a}_l \ket{0}  \label{eq:F1F2_term_line1} \\
    &= \frac{1}{4}\sum_{i,j,k,l}\hat{I}_{ij}\hat{a}_{i}\hat{a}_{j}\hat{A}_{i}\hat{A}_{j}(\delta_{ik}\delta_{jl} + \delta_{il} \delta_{jk}) \ket{0} \label{eq:F1F2_term_line2}  \\
    &= \ket{0, 1},
\end{align}
where the first step uses the identities $\hat{a}_i\tilde{a}_i = \hat{a}_i$ and $[\bar{A}_i, \hat{A}_j] = \delta_{ij} \hat{A}_i$. We will soon show more generally that
\begin{equation}
	\frac{\mathbb{F}_2^{p}}{p!} \frac{\mathbb{F}_1^{q}}{q!} \ket{0} = \left\{ \begin{array}{cl} \ket{q-2p, p} & \textrm{if $q \geq 2p$}, \\ 0 & \textrm{otherwise}. \end{array} \right.
	\label{eq:general_homodimer_factory_operator_product_on_vacuum}
\end{equation}
Summing this over all $p$ and $q$ establishes the $\ket{\rm sum}$ state in \eq{eq:factory_form_of_homodimer_sum_vector}.

The sum of states for complexes generated in any system can thus be specified by a vector of operators $\vec{\mathbb{F}} = (\mathbb{F}_1, \ldots, \mathbb{F}_L)^\top$ via 
\begin{equation}
    \sumket = e^{\mathbb{F}_{L}} \ldots e^{\mathbb{F}_{1}}\ket{0}. \label{eq:sumket_from_factory}
\end{equation}
We call this vector the ``factory.'' We emphasize that the order of the operators within the factory is important, as these operators, unlike gallery operators, do not generally commute. 

Using the factory instead of the gallery to define the set of possible complexes in a system can have an important advantage: the factory often comprises far fewer operators than the gallery. This is not the case for the homodimer system, but it is so for the homopolymer system presented in Section \ref{sec:homopol}. 

There is a disadvantage, however, to defining a system using the factory: one loses access to the generating function. One can, of course define a coherent state analogous to $\ket{\vec{z}}$  via
\begin{equation}
    \ket{\vec{x}} = e^{x_L \mathbb{F}_L} \cdots e^{x_1 \mathbb{F}_1} \ket{0}, \label{eq:coherent_state_x}
\end{equation}
where $\vec{x} = (x_1, \ldots, x_L)^\top$. It is questionable, however, how useful the corresponding generating function $\braket{\vec{x} | \psi}$ is for analysis. As we will see, each term in the expansion of \eq{eq:coherent_state_x} can yield multiple distinct mixtures of complexes. One thus generally cannot read off the macrostate distribution $P(\vec{n})$ from the expansion of $\braket{\vec{x} | \psi}$.

\subsection{Wick's theorem}

The algebraic manipulations needed to show \eq{eq:general_homodimer_factory_operator_product_on_vacuum} become unwieldy as $p$ and $q$ become large. Wick's theorem, a foundational result in quantum field theory, makes these calculations significantly more straightforward by providing a systematic procedure for reordering operators in a multi-operator product. 

To see how Wick's theorem can be applied to our formalism, define the compound operators $\bar{A}^a_i = \hat{a}_i \bar{A}_i$ and $\hat{A}^a_i = \hat{a}_i \hat{A}_i$. Ignoring the interaction field $I$ for the moment, each term in the expansion of the left-hand side of \eq{eq:general_homodimer_factory_operator_product_on_vacuum} has the form
\begin{equation}
    \bar{A}^a_{i_1} \cdots \bar{A}^a_{i_{2q}} \hat{M}_{j_1} \cdots \hat{M}_{j_p} \ket{0}. \label{eq:factory_term}
\end{equation}
Note that all compound presence operators appear to the left of all creation operators. We refer to this as ``productive ordering.'' Given an operator product $X_1 X_2 \cdots X_n$, we denote its productive ordering by $\mathcal{P}(X_1 X_2 \cdots X_n)$. Each term in the Taylor expansion of the factory representation is productive ordered because the instructions for assembling each complex are applied after the instructions for creating its components. In contrast, each term in the expansion of the gallery representation in \eq{eq:gallery_representation_of_expansion} has the form \begin{equation}
    \hat{M}_{i_1} \cdots \hat{A}_{i_m} \hat{M}^a_{j_1} \cdots \hat{A}^a_{j_{2d}}  \ket{0}. \label{eq:gallery_term}
\end{equation}
The key difference from \eq{eq:factory_term} is that this term contains only creation operators, all of which commute. 

Every term of the form in \eq{eq:factory_term} is in fact equal to a sum of terms having the form in \eq{eq:gallery_term} with $m = q-2p$ and $d=p$. To transform the former to the latter, we iteratively apply the exchange rule
\begin{align}
    \bar{A}^a_i \hat{M}_j &= \hat{M}_j \bar{A}^a_i + \delta_{ij} \hat{A}^a_i \label{eq:wick_exchange}
\end{align}
until no $\bar{A}^a_i$ operators appear to the left of any $\hat{M}_j$ operators. Each application of the exchange rule adds another term to the expansion. The result is a sum of operator products such that all  $\bar{A}^a_i$ in each product appear to the right of all $\hat{M}_i$ and $\hat{A}^a_i$. Such products are said to be ``normally ordered.'' More generally,  an operator product  is normally ordered if all presence operators appear to the right of all creation operators. The normally ordered form of an operator product $X_1 X_2 \cdots X_n$ is denoted by $\mathcal{N}(X_1 X_2 \cdots X_n)$. Normally ordered products are useful because any such products containing presence operators vanish when applied to the vacuum state. 

Wick's theorem provides an equality between productive ordered and normally ordered operator products. To state Wick's theorem, we define a ``contraction'' between two operators $X_{i}$ and $X_{j}$ to be
\begin{equation}
    \wick{\c1{X_i} \c1{X_j}} = \mathcal{P}(X_i X_j) - \mathcal{N}(X_i X_j).
\end{equation}
The contraction of two specific operators within a larger product $X_1 X_2 \cdots X_n$ removes these operators from the product and replaces them with their contraction, i.e.,
\begin{equation}
    \wick{X_1 X_2 \cdots \c1{X_i} \cdots \c1{X_j} \cdots X_n} =  X_1 X_2 \dots X_n (\wick{\c1{X_i}\c1{X_j}}).
\end{equation}

A key assumption of Wick's theorem is that the contraction of any two operators in a product is ``central,'' i.e., it commutes with all other operators in the product. For the homodimer algebra, the only contractions needed to transform \eq{eq:factory_term} to \eq{eq:gallery_term} are of the form
\begin{align}
    \wick{\c1{\bar{A}^a_i} \c1{\hat{M}_j}} = \bar{A}^a_i \hat{M}_j - \hat{M}_j \bar{A}^a_i = [\bar{A}^a_i, \hat{M}_j] = \delta_{ij} \hat{A}^a_j.
\end{align}
These contractions are indeed central, i.e., $[\hat{A}^a_j, \hat{M}_j] = [\hat{A}^a_j, \bar{A}^a_i] = 0$.

Applied to our context, Wick's theorem states that any productive ordered operator product is equal to the sum of all possible normally ordered contractions:
\begin{widetext}
\begin{align}
    \mathcal{P}(X_1 X_2 X_3 X_4 \cdots ) = &\:\mathcal{N}(X_1 X_2 X_3 X_4  \cdots ) + \nonumber \\
    & \: \mathcal{N}(\wick{\c1{X_1} \c1{X_2} X_3 X_4 \cdots}) + \mathcal{N}(\wick{\c1{X_1} X_2 \c1{X_3} X_4 \cdots}) + \mathcal{N}(\wick{X_1 \c1{X_2} \c1{X_3} X_4 \cdots}) + \cdots \nonumber \\
    & \: \mathcal{N}(\wick{\c1{X_1} \c1{X_2} \c1{X_3} \c1{X_4} \cdots}) + \mathcal{N}(\wick{\c1{X_1} \c2{X_2} \c1{X_3} \c2{X_4} \cdots}) + \mathcal{N}(\wick{\c1{X_1} \c2{X_2} \c2{X_3} \c1{X_4} \cdots}) + \cdots \nonumber\\
    & \: \vdots \\
    = &\:\sum_{\mathrm{all~contractions}~\mathcal{C}} \mathcal{N}(\mathcal{C}(X_1 X_2 \ldots X_n))
\end{align}
For example, applying Wick's theorem to the right-hand side of \eq{eq:F1F2_term_line1} gives
\begin{align}
    \bar{A}^a_i \bar{A}^a_j \hat{M}_k \hat{M}_k 
    = &\:\mathcal{N}(\bar{A}^a_i \bar{A}^a_j \hat{M}_k \hat{M}_k) + \nonumber \\
    & \: \mathcal{N}(\wick{ \c1{\bar{A}^a_i} {\bar{A}^a_j} \c1{\hat{M}_k} {\hat{M}_k} }) +
        \mathcal{N}(\wick{ \c1{\bar{A}^a_i} {\bar{A}^a_j} {\hat{M}_k} \c1{\hat{M}_k} }) +
        \mathcal{N}(\wick{ {\bar{A}^a_i} \c1{\bar{A}^a_j} \c1{\hat{M}_k} {\hat{M}_k} }) +
        \mathcal{N}(\wick{ {\bar{A}^a_i} \c1{\bar{A}^a_j} {\hat{M}_k} \c1{\hat{M}_k} }) + \nonumber \\
    & \: \mathcal{N}(\wick{ \c1{\bar{A}^a_i} \c2{\bar{A}^a_j} \c1{\hat{M}_k} \c2{\hat{M}_k} }) + 
        \mathcal{N}(\wick{ \c1{\bar{A}^a_i} \c2{\bar{A}^a_j} \c2{\hat{M}_k} \c1{\hat{M}_k} })~ \\ 
    = &\:\hat{M}_k \hat{M}_k \bar{A}^a_i \bar{A}^a_j + \nonumber \\
    & \: \delta_{ik} \hat{A}^a_i \hat{M}_k \bar{A}^a_j + 
        \delta_{il} \hat{A}^a_i \hat{M}_k \bar{A}^a_j + 
        \delta_{jk} \hat{A}^a_j \hat{M}_k \bar{A}^a_i +
        \delta_{jl} \hat{A}^a_j \hat{M}_k \bar{A}^a_i +  \nonumber \\
    & \: \delta_{ik} \delta_{jl} \hat{A}^a_i \hat{A}^a_j + 
        \delta_{il} \delta_{jk} \hat{A}^a_i \hat{A}^a_j~. \label{eq:last_two_terms}
\end{align}
\end{widetext}
When applied to the vacuum state, only the last two terms in \eq{eq:last_two_terms} survive, thus yielding the expression in \eq{eq:F1F2_term_line2},
\begin{eqnarray}
    \bar{A}^a_i \bar{A}^a_j \hat{M}_k \hat{M}_k \ket{0}
    &=& (\delta_{ik} \delta_{jl} + \delta_{il} \delta_{jk}) \hat{A}^a_k \hat{A}^a_l \ket{0}.
\end{eqnarray}
More generally, Wick's Theorem allows us to transform terms in the expansion of the factory expression for the sum vector (\eq{eq:sumket_from_factory}) into a sum of terms in the expansion of the gallery expression for the sum vector (\eq{eq:gallery_sum_state}).

\subsection{Formal diagrams} \label{sec:homodimer-diagrams}

\begin{figure}[b]
    \centering
    \boxed{\includegraphics[width=0.97\columnwidth]{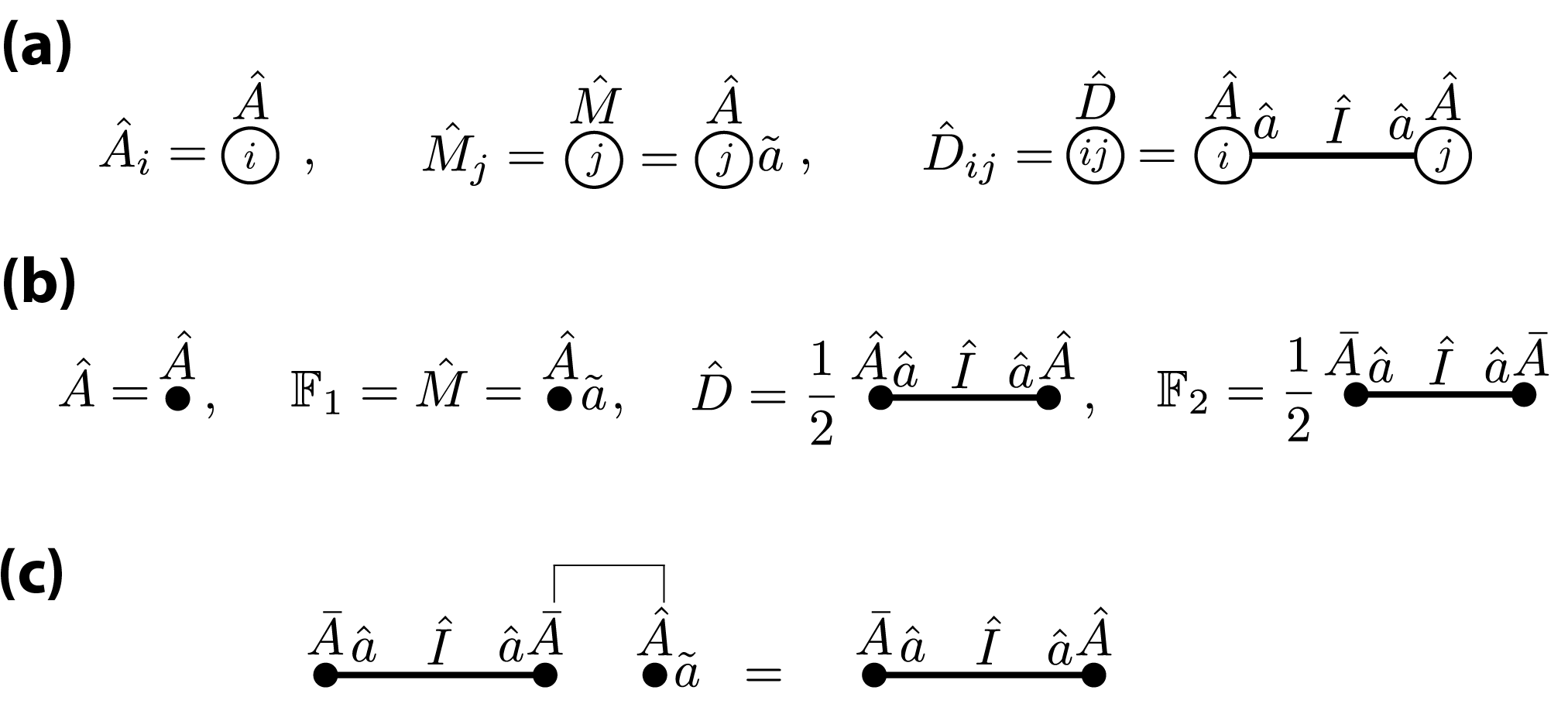}}
    \caption{Diagrammatic notation for operator products and sums thereof over internal indices. (a) Examples of simple and compound mode operators. (b) Examples of simple and compound field operators. (c) Wick contractions relevant to the homodimer system.}
    \label{fig:first_diagrams}
\end{figure}

We now introduce diagrammatic methods that aid in computations involving Fock space operators. Each diagram indicates an operator product or sums of such products over internal states. \fig{fig:first_diagrams}{} shows several examples. The indices of mode operators are written as index names inside open dots [\fig{fig:first_diagrams}{(a)}]. Mode operators are indicated by the decorated operator name written next to their respective dots. Multiple operator names written next to the same dot indicate that those operators share the same index. Modes that have two indices are written next to lines that connect the two dots representing these indices. A closed dot indicates summation over the corresponding index. Field operators are thus distinguished from mode operators through the use of closed rather than open dots. Symmetry factors are also kept explicit [\fig{fig:first_diagrams}{(b)}]. 

\begin{figure}[t]
    \centering
    \fbox{\includegraphics[width=0.97\columnwidth]{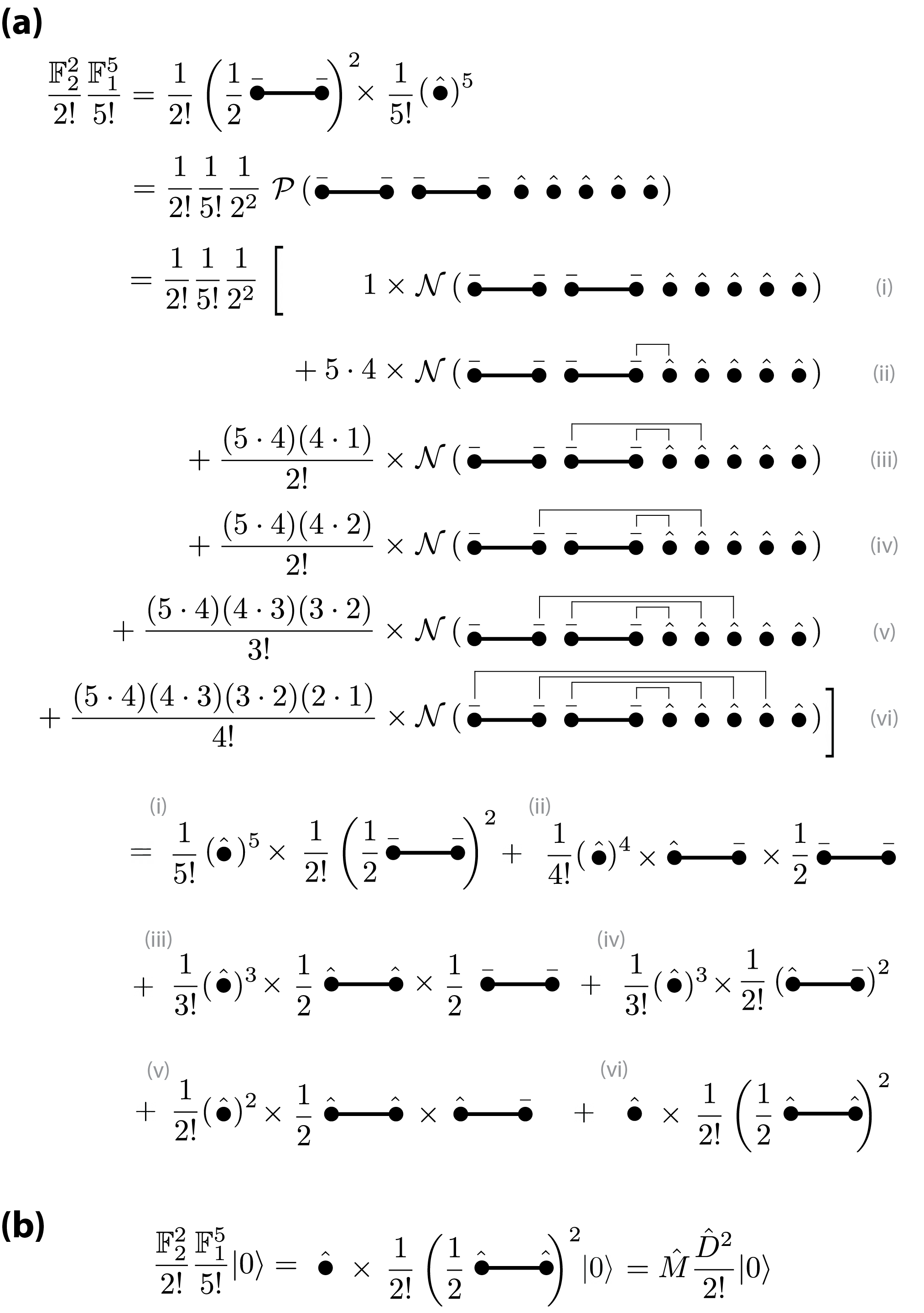}}
    \caption{Diagrams facilitate algebraic calculations. (a) Evaluation of the $p=2$, $q=5$ term of \eq{eq:general_homodimer_factory_operator_product_on_vacuum} in terms of normally ordered operator products. (b) Result of the computation in panel (a) applied to the vacuum state.}
    \label{fig:example_diagram_computation}
\end{figure}

This diagrammatic notation is helpful in computations involving Wick contractions; we demonstrate this by deriving \eq{eq:general_homodimer_factory_operator_product_on_vacuum}. The effect of each Wick contraction is illustrated in \fig{fig:first_diagrams}{(c)}: contracting an $\bar{A}^a$ operator (part of $\mathbb{F}_2$) with $\mathbb{F}_1 = \hat{M}$ eliminates the $\mathbb{F}_1$ and replaces the $\bar{A}^a$ in $\mathbb{F}_2$ with an $\hat{A}^a$. To avoid unnecessary notation going forward, we represent this operation using the same diagrams but showing only the decorations on the $A$ operators. Now consider the left-hand side of \eq{eq:general_homodimer_factory_operator_product_on_vacuum} with $p=2$ and $q=5$ [\fig{fig:example_diagram_computation}{(a)}, line 1]. Because the two $\mathbb{F}_2$ operators are applied after the five $\mathbb{F}_1$ operators, this product is productive ordered [\fig{fig:example_diagram_computation}{(a)}, line 2]. Next we use Wick's theorem to convert this to a sum of normally ordered products [\fig{fig:example_diagram_computation}{(a)}, lines 3-8]. We also evaluate the combinatorial coefficients that arise due to distinct contractions producing topologically identical products. Consider, for example, the coefficient for term (iii). For the first contraction, there are five choices of $\hat{M}$ and four choices of $\bar{A}^a$. For the second contraction, there are four remaining choices of $\hat{A}$ but only one possible choice for $\bar{A}^a$ -- that which is linked with the first $\bar{A}^a$ through an $I$-field. Since interchanging the order in which the contractions are performed does not change the result, we divide the result by two. This yields a combinatorial coefficient of $(5\cdot 4) \times (4 \cdot 1) / 2!$. The combinatorial coefficients for the other terms follow similarly. 

Each normally ordered term in lines 3-8 yields one of the resulting operator products shown in lines 9-11, as indicated. We leave it to the reader to check that the combinatorial coefficients computed in lines 3-8 do in fact match those shown in lines 9-11, which are as expected based on symmetry considerations. Note in particular that all terms except term (vi) contain $\bar{A}^a$ operators. Consequently, only term (vi) survives when applying this result to the vacuum state [\fig{fig:example_diagram_computation}{(b)}]. 

\begin{figure}[t]
    \centering
    \fbox{\includegraphics[width=0.97\columnwidth]{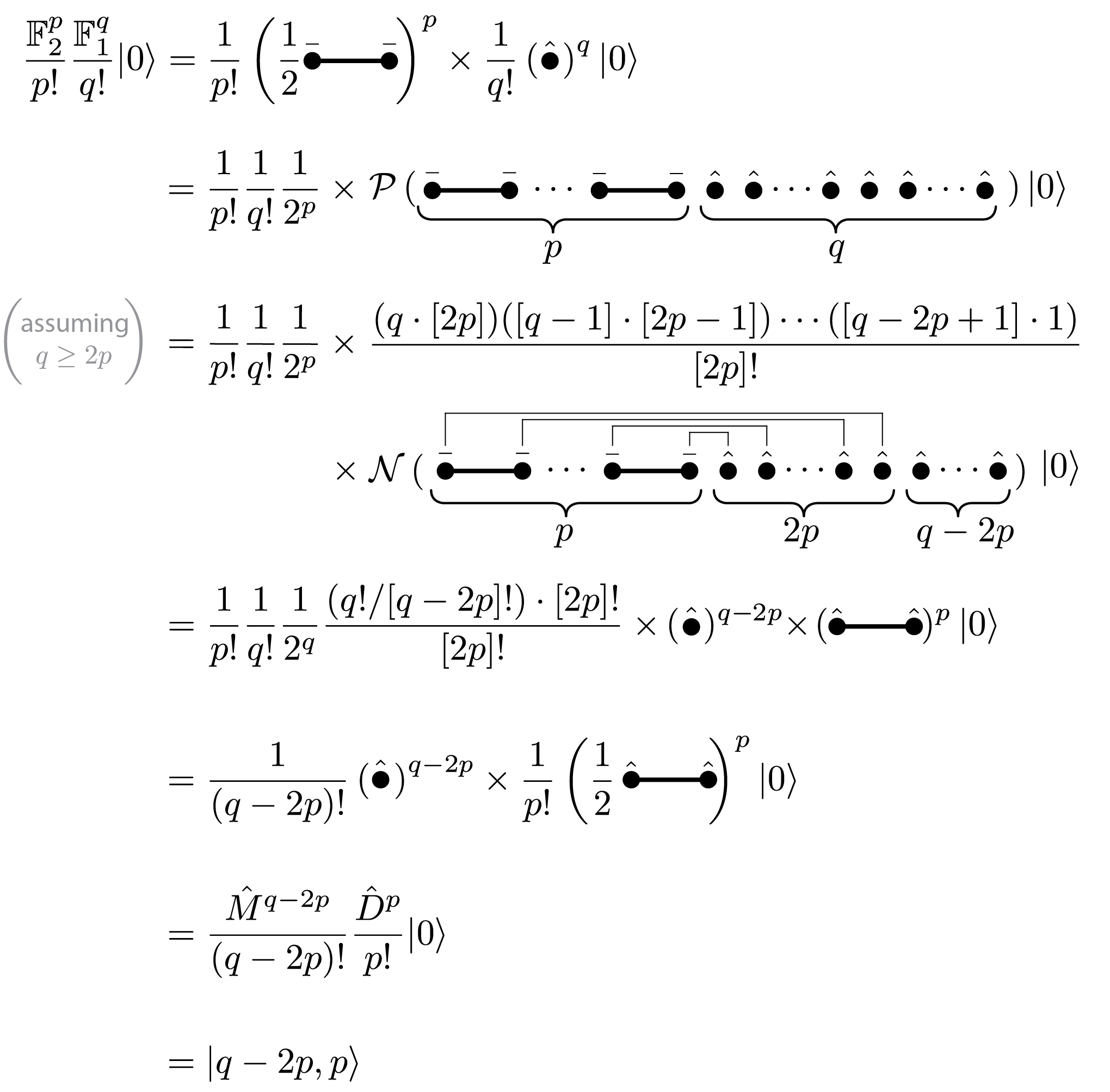}}
    \caption{Diagrammatic proof of the factory/gallery equivalence for the homodimer system.}
    \label{fig:general_homodimer}
\end{figure}

We are now in a position to evaluate the left-hand side of \eq{eq:general_homodimer_factory_operator_product_on_vacuum} for general values of $p$ and $q$ (\fig{fig:general_homodimer}{}). If $q < 2p$, then $\mathbb{F}_1^q$ does not supply enough $\hat{M}$ operators to contract all the $\bar{A}^a$ operators supplied by $\mathbb{F}_2^p$. The expression therefore vanishes. If $q \geq 2p$, however, there are $q!/(q-2p)!$ ways to contract all the $\bar{A}^a$ with all the $\hat{M}$, thereby leaving $p$ copies of $\hat{D}$ and $q$ copies of $\hat{M}$. The resulting combinatorial factor replaces the $1/q!$ with $1/(q-2p)!$, thus providing the factors needed to correct the redundancies in $\hat{D}^p$ and $\hat{M}^q$. This completes the derivation of \eq{eq:general_homodimer_factory_operator_product_on_vacuum}, and thus proof of the factory/gallery equivalence for the homodimer system.

\section{\label{sec:homopol}Homopolymer in equilibrium}
\begin{figure}[b]
    \centering
    \fbox{\includegraphics[width=0.97\columnwidth]{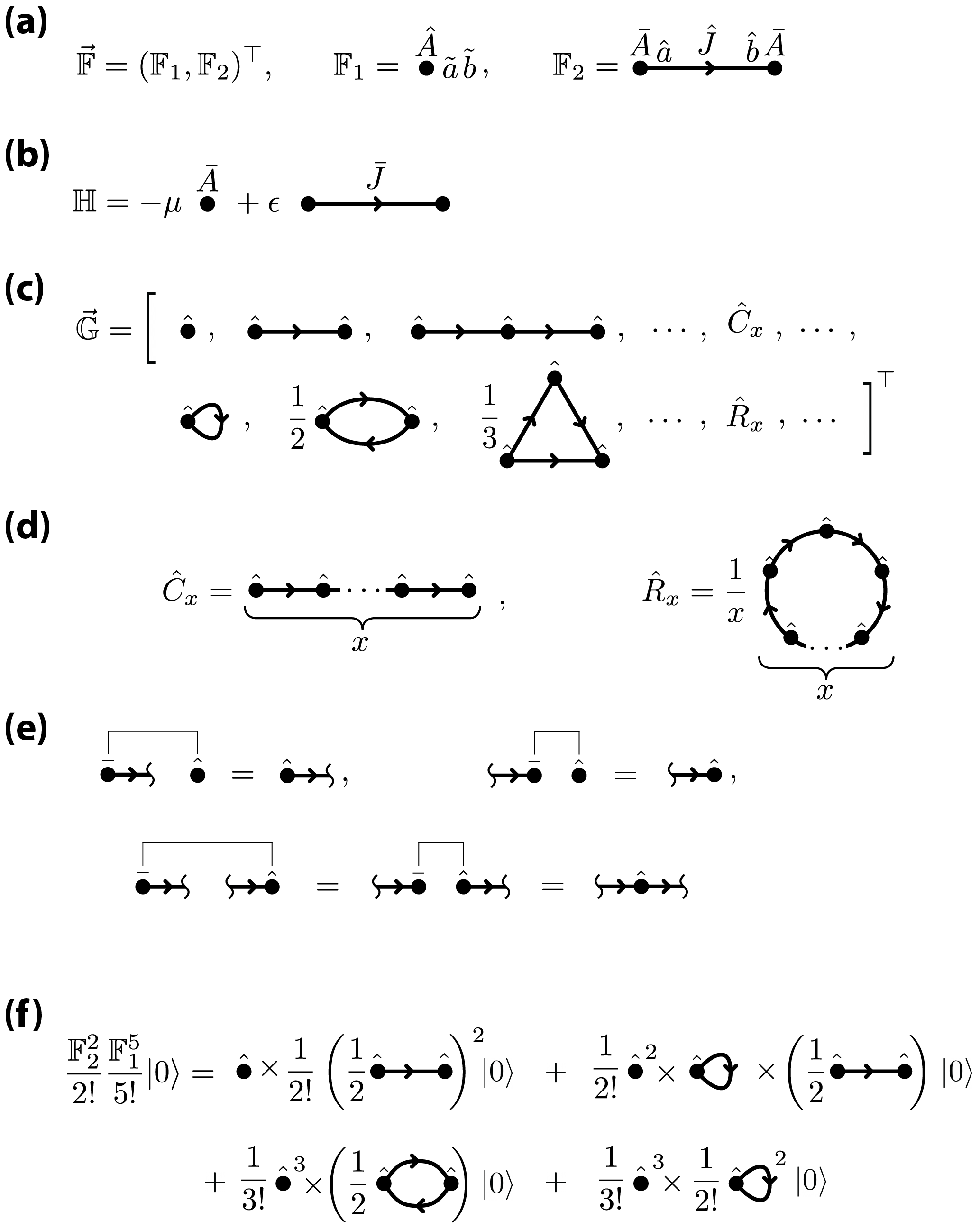}}
    \caption{Diagrammatic specification of the homopolymer system. (a) The factory. (b) The Hamiltonian. (c) The resulting gallery. (d) The $\hat{C}_x$ and $\hat{R}_x$ comprising the gallery, and (e) the four contraction rules for the system. For conciseness, panels (c-e) hide all operator names except for the decorators on the $A$ operators. (f) An example algebraic calculation carried out using the commutation rules in panel (e).}
    \label{fig:homopolymer_specification}
\end{figure}

We now turn to a system that is far simpler to define in a rule-based manner than in a species-based manner. Consider a factory comprising two operators:
\begin{equation}
	\mathbb{F}_1 = \sum_i \hat{A}_i \tilde{a}_i \tilde{b}_i,~~~\mathbb{F}_2 = \sum_{i,j} \bar{A}_i \bar{A}_j \hat{a}_i \hat{b}_j \hat{J}_{ij}.
\end{equation}
This is similar to the homodimer factory, but $\mathbb{F}_2$ differs in that it forms an asymmetric (rather than symmetric) bond between two $A$ particles. Specifically, the summand in $\mathbb{F}_2$ occupies a site $a_i$ on the $A_i$ monomer, a site $b_j$ on the $A_j$ monomer, and forms a bond $J_{ij}$ between them.  $\mathbb{F}_2$ is not multiplied by a symmetry factor because $J_{ij} \neq J_{ji}$. These factory operators are represented graphically in \fig{fig:homopolymer_specification}{(a)}. We define a rule-based Hamiltonian for this system in the familiar way:
\begin{equation}
	\mathbb{H} = - \mu \sum_i \bar{A}_i + \epsilon \sum_{i,j}\bar{J}_{ij},
\end{equation}
as represented in \fig{fig:homopolymer_specification}{(b)}. The resulting gallery [\fig{fig:homopolymer_specification}{(c)}] is far more complex than that of the homodimer: it comprises creation operators for polymer chains and polymer rings of all lengths. Here, $x$-chains and $x$-rings are created by the operators 
\begin{align}
	\hat{C}_x &= \sum_{i_1, \ldots, i_x}  \hat{A}_{i_1} \cdots \hat{A}_{i_x} \hat{a}_{i_1} \cdots \hat{a}_{i_{x-1}} \\
		& ~~~~~~~~~~~~~ \times  \hat{b}_{i_2} \cdots \hat{b}_{i_x} \hat{J}_{i_1 i_2} \cdots \hat{J}_{i_{x-1} i_x}, \nonumber  \\
	\hat{R}_x &=  \frac{1}{x} \sum_{i_1, \ldots, i_x}  \hat{A}_{i_1} \cdots \hat{A}_{i_x} \hat{a}_{i_1} \cdots \hat{a}_{i_x} \label{eq:factor_of_1_over_x} \\
	 	& ~~~~~~~~~~~~~ \times \hat{b}_{i_1} \cdots \hat{b}_{i_x} \hat{J}_{i_1 i_2} \cdots \hat{J}_{i_{x-1} i_x} \hat{J}_{i_{x} i_1}.\nonumber
\end{align}
These operators are more clearly expressed in diagrammatic notation [\fig{fig:homopolymer_specification}{(d)}]. Note the factor of $1/x$ in \eq{eq:factor_of_1_over_x}; this is needed to compensate for redundancy in the sum over internal indices that results from $x$-rings having rotational symmetry.

The equivalent species-based Hamiltonian has an infinite number of terms, each with its own bare chemical potential: 
\begin{equation}
	\mathbb{H} \simeq -\sum_{x=1}^\infty \left( \mu_{C_x}\, \bar{C}_x +  \mu_{R_x}\, \bar{R}_x \right),
\end{equation}
where $\mu_{C_x} = x \mu - (x-1)\epsilon$ and $\mu_{R_x} = x \mu - x\epsilon$ are the bare chemical potentials for $x$-chains and $x$-rings. The corresponding number of complex-specific microstates are $N_{C_{x}} = N^x$ and $N_{R_{x}} = \frac{1}{x}N^x$. Putting these together, we obtain the effective chemical potentials of each species:
\begin{align}
    \mu'_{C_x} &= x(\mu - \epsilon) + \epsilon - kT \log \frac{N^x}{V}, \nonumber \\
    \mu'_{R_x} &= x(\mu - \epsilon) - kT \log \frac{N^x}{xV}. \label{eq:homopolymer_effective_chemical_potentials}
\end{align}
Unlike in the homopolymer system, it is not possible to renormalize $\mu$ and $\epsilon$ so that all $\mu'_{C_x}$ and $\mu'_{R_x}$ are independent of volume. Keeping $\mu'_{C_1}$ independent of $N$ and $V$ requires fixing the value of $\mu' = \mu - k_B T \log \frac{N}{V}$ as in the monomer and homodimer systems. Keeping all other $\mu'_{C_x}$ independent of $N$ and $V$ then requires fixing $\epsilon' = \epsilon + k_B T \log V$. The effective chemical potentials for all species thus become
\begin{align}
    \mu'_{C_x} &= x(\mu' - \epsilon') + \epsilon', \nonumber \\
    \mu'_{R_x} &= x(\mu' - \epsilon') - k_B T \log x - k_B T \log V. \label{eq:homopolymer_effective_chemical_potentials_renorm}
\end{align}
Constraining the concentrations of all $C_x$ species independent of $V$ therefore requires that the concentration of all $R_x$ species scale as $V^{-1}$. We note, however, that the converse is not possible, i.e., one cannot choose a definition for $\mu'$ and $\epsilon'$ so that the concentrations of all ring species are independent of $V$. 

We therefore conclude that, for the homopolymer system to behave sensibly in the $V \to \infty$ limit, the concentrations of all chain polymers must remain fixed whereas the concentrations of all ring polymers must vanish. This makes sense: as $V$ increases, the number of free ends with which the free end of an $x$-chain can interact increases in proportion to $V$. To preserve the concentrations of all $x$-chains, $e^{-\beta \epsilon}$ must scale as $V^{-1}$. The probability of one free end of an $x$-chain interacting with the other free end of the same polymer will thus scale as $V^{-1}$, and the concentration of all ring polymers will also scale as $V^{-1}$. 

Given \eq{eq:homopolymer_effective_chemical_potentials_renorm}, defining $\eta = e^{\beta(\mu' - \epsilon')}$ and computing
\begin{align}
    \log Z = \sum_{x=0}^\infty e^{\beta \mu_{C_x}} + \sum_{x=0}^\infty e^{\beta \mu_{R_x}},
\end{align} 
we find that the log partition function density of the system is, for $0 < \eta < 1$,
\begin{align}
	\frac{\log Z}{V} = \frac{e^{\beta \epsilon'}}{1 - \eta} - \frac{\log ( 1 - \eta )}{V}.
\end{align}
The left and right terms of the resulting expression are the respective contributions from chains and rings. The $V^{-1}$ scaling of the second term reflects the vanishing of ring species as $V\to\infty$. The remaining chain contribution diverges as $\eta \to 1$ from below. In this limit, the concentration of $A$ particles diverges as
\begin{align}
	\frac{\braket{\bar{A}}}{V} = \frac{1}{\beta} \frac{\partial}{\partial \mu'} \frac{\log Z}{V} \approx \frac{e^{\beta \epsilon'}}{\delta^2}. \label{eq:homopolymer_A_cond_div}
\end{align}
Since the concentration of each chain species $C_x$ is given by $e^{\beta \mu'_{C_x}} = e^{\beta \epsilon'} \eta^x$, the distribution of chain lengths is distributed exponentially with decay rate $\log \eta$, and flattens out as $\eta \to 1$ from below. Defining $\delta = 1 - \eta$, the mean and variance of these chain lengths diverge as
\begin{align}
    \braket{x} = \frac{\eta}{1\!-\!\eta} \approx \frac{1}{\delta},~~\textrm{var}(x) = \frac{\eta}{(1\!-\!\eta)^2} \approx \frac{1}{\delta^2}.
\end{align}

To show the equivalence of the factory and gallery representations of the homopolymer system, we again invoke Wick's theorem. In this case, however, the allowable contractions are more complex. Paralleling the analysis for the homodimer, we define the compound operators
\begin{align}
    \hat{M}_i &= \hat{A}_i \tilde{a}_i \tilde{b}_i,~~~\hat{A}_i^a = \hat{A}_i \hat{a}_i \tilde{b}_i, \\
    \hat{A}_i^b &= \hat{A}_i \tilde{a}_i \hat{b}_i,~~~\hat{A}_i^{ab} = \hat{A}_i \hat{a}_i \hat{b}_i, \\
    \bar{A}_i^a &= \bar{A}_i \hat{a}_i \tilde{b}_i,~~~\bar{A}_i^b = \bar{A}_i \tilde{a}_i \hat{b}_i.
\end{align}
These six operators obey the contraction rules
\begin{align}
	\wick{\c1{\bar{A}^a_i} \c1{\hat{M}_j}} &= \delta_{ij} \hat{A}^a_i, ~~~~\wick{\c1{\bar{A}^b_i} \c1{\hat{M}_j}} = \delta_{ij} \hat{A}^b_i \\
	\wick{\c1{\bar{A}^a_i} \c1{\hat{A}^b_j}} &= \wick{\c1{\bar{A}^b_i} \c1{\hat{A}^a_j}} = \delta_{ij} \hat{A}^{ab}_i. 
\end{align}
These rules are shown diagrammatically in \fig{fig:homopolymer_specification}{(e)}. The reader may notice that the first two contraction products are not central, and thus violate an assumption of Wick's theorem. We find, however, that Wick's theorem still holds if we allow contraction products to participate in additional contractions. Fully contracted operator products therefore have all $\bar{A}_i^a$ and $\bar{A}_i^b$ operators participating in one contraction each, whereas each $\hat{A}$ operator may participate in zero, one, or two contractions.

We can use the above contraction rules to derive the complexes generated from any given term in $\ket{\rm sum} = e^{\mathbb{F}_{2}}e^{\mathbb{F}_{1}}\ket{0}$. \fig{fig:homopolymer_specification}{(f)} shows the result for one such term. A generalized version of this computation is used in Appendix \ref{app:FGequivpol} to prove the factory/gallery equivalence for the homopolymer system. 

\section{\label{sec:noneq}Nonequilibrium systems}
\subsection{Species-based formalism}

We now turn to the problem of determining the macrostate master equation [\eq{eq:macrostate_master_equation}] given a rule-based microstate master equation [\eq{eq:microstate_master_equation}]. We start, however, by investigating how to specify and analyze a more standard species-based microstate master equation. 

Species-based master equations are built from individual reactions, each of which annihilates a specified set of complexes and creates a new set in their place. Suppose there are $H$ distinct species-specific reactions. The transition operator in \eq{eq:microstate_master_equation} will then have the form 
\begin{equation}
    \mathbb{W} = \sum_{h=1}^H r_h \left( \mathbb{Q}_h - \grave{\mathbb{Q}}_h \right), \label{eq:species_based_transition}
\end{equation}
where $r_h$ is the rate at which reaction $h$ occurs, $\mathbb{Q}_h$ is a ``reaction operator'' that effects this reaction when applied to a macrostate $\ket{\vec{n}}$, and $\grave{\mathbb{Q}}$ is a corresponding ``depletion operator.'' Each reaction operator has the form
\begin{equation}
	\mathbb{Q} = \prod_{k=1}^K \frac{\hat{G}_k^{p_k}}{p_k!} \frac{\check{G}_k^{q_k}}{q_k!}, \label{eq:Q}
\end{equation}
where  $\vec{q} = (q_1, \ldots, q_K)$ is an abundance vector that describes the reactants and $\vec{p} = (p_1, \ldots, p_K)$ is a vector that describes the products. For the sake of simplicity we assume that $\vec{p}$ and $\vec{q}$ do not overlap (i.e., $p_k = 0$ and/or $q_k = 0$ for every $k$). Applying the conjugate of this operator to the macrostate, one finds that
\begin{align}
    \mathbb{Q}^\dagger \ket{\vec{n}} &= N_{\vec{p}}\, \Omega(\vec{n}, \vec{q}, \vec{p}) \ket{\vec{n}+\vec{q}-\vec{p}}, \label{eq:Q_on_n} 
\end{align}
where \begin{align}
	\Omega(\vec{n}, \vec{q}, \vec{p}) &= \prod_k {n_k + q_k - p_k \choose q_k} 1(n_k \geq p_k)
\end{align}
is a coefficient that depends on $\vec{n}$, $\vec{p}$, and $\vec{q}$, but not otherwise on the details of the reaction, and 
\begin{equation}
	N_{\vec{p}} = \prod_k {N_k \choose p_k} \approx \prod_k \frac{N_k^{p_k}}{p_k!}
\end{equation}
is the number of distinct product microstates created when $\mathbb{Q}$ is applied to a single reactant microstate. 

In our formalism, the depletion operator $\grave{\mathbb{O}}$ corresponding to any reaction operator $\mathbb{O}$ is given in terms of the microstate-specific components via
\begin{align}
    \mathbb{O} = \sum_{\mathcal{I},\mathcal{J}} o_{\mathcal{I}\mathcal{J}} \ket{\mathcal{J}}\bra{\mathcal{I}} ~~\Rightarrow ~~
    \grave{\mathbb{O}} = \sum_{\mathcal{I},\mathcal{J}} o_{\mathcal{I}\mathcal{J}} \ket{\mathcal{I}}\bra{\mathcal{I}},  \label{eq:depletion_operator_def}
\end{align}
where $\mathcal{I}$ and $\mathcal{J}$ index all microstates of the system and $o_{\mathcal{I}\mathcal{J}}$ is the rate at which $\ket{\mathcal{I}}$ is transformed into $\ket{\mathcal{J}}$. Note that, by this definition, all depletion operators are self-conjugate. In Appendix \ref{app:depderiv} we show that \eq{eq:depletion_operator_def}, together with the assumption of non-overlapping $\vec{p}$ and $\vec{q}$, leads to  a depletion operator corresponding to $\mathbb{Q}$ of
\begin{equation}
	\grave{\mathbb{Q}} = \prod_k { \tilde{G}_k \choose p_k} {\bar{G}_k \choose q_k}. \label{eq:Qgrav}
\end{equation}
Applying $\grave{\mathbb{Q}}^\dagger = \grave{\mathbb{Q}}$ to the macrostate then gives
\begin{align}
    \grave{\mathbb{Q}} \ket{\vec{n}} &= N_{\vec{p}}\, \Omega(\vec{n}, \vec{q}, \vec{q}) \ket{\vec{n}}. \label{eq:Qdep_on_n}
\end{align}
 Adding back the subscript $h$ on $\vec{p}$ and $\vec{q}$ and using \eq{eq:J_vec_def}, we obtain an expression for the flux projector:
\begin{align}
    \ket{J(\vec{n})} 
    & = \sum_{h=1}^H r_h N_{\vec{p}_h} \{ \Omega(\vec{n}, \vec{q}_h, \vec{p}_h) \ket{\vec{n} + \vec{q}_h - \vec{p}_h} \nonumber \\ 
    & ~~~~~~~~~~~~~~~- \Omega(\vec{n}, \vec{q}_h, \vec{q}_h) \ket{\vec{n}} \}.
    \label{eq:speciesME}
\end{align}

The difficulty with this species-based formulation is that, in systems that admit multi-particle complexes, the form of the transition operator in \eq{eq:species_based_transition} does not reflect the underlying simplicity of the system. Rather, the rates $r_h$ and reaction operators $\mathbb{Q}_h$ are derived quantities that follow from an (often much smaller) set of informally stated rules. Furthermore, even manually specifying the right-hand side of \eq{eq:species_based_transition} can be tricky: a small number of rules can lead to a very large (or even infinite) number of reactions $H$, and both $r_h$ and $\mathbb{Q}_h$ can depend  in nontrivial ways on the elemental parameters that govern those rules. We now show how our formalism addresses this problem by enabling the rule-based definition of $\mathbb{W}$.

\subsection{Rule-based formalism}

We specify the transition operator in a rule-based manner as follows. Suppose we have a system defined by $L$ reaction rules. For each rule $l$, we specify a rate $r_l$ and a ``reaction rule operator'' $\mathbb{R}_l$. The transition operator is then given by
\begin{equation}
    \mathbb{W} = \sum_{l=1}^L r_l \left( \mathbb{R}_l - \grave{\mathbb{R}}_l \right), \label{eq:W_from_R}
\end{equation}
where $\grave{\mathbb{R}}_l$ is the depletion operator corresponding to $\mathbb{R}_l$. 

Suppose a rule operator $\mathbb{R}$ is able to drive $M$ different species-specific reactions. For each reaction $m$, let $\vec{q}_m$ denote the number of reactant species and $\vec{p}_m$ the number of product species. We find that
\begin{align}
    \mathbb{R}^\dagger \ket{\vec{n}} &= \sum_{m=1}^M \sigma_m \Omega(\vec{n}, \vec{q}_m, \vec{p}_m) \ket{\vec{n}+\vec{q}_m-\vec{p}_m}, \label{eq:R_on_n}
\end{align}
where $\sigma_m$ is the number of distinct product microstates that can result from each reactant microstate in an $m$-type reaction. The depletion operator $\grave{\mathbb{R}}$ follows from the rule operator $\mathbb{R}$ using \eq{eq:depletion_operator_def}. Applying $\grave{\mathbb{R}}^\dagger = \grave{\mathbb{R}}$ to $\ket{\vec{n}}$,
\begin{align}
    \grave{\mathbb{R}} \ket{\vec{n}} = \sum_{m=1}^M N_{\vec{p}_m}\, \Omega(\vec{n}, \vec{q}_m, \vec{q}_m) \ket{\vec{n}}. \label{eq:Rdep_on_n} 
\end{align}
Adding back the $l$ indices, we obtain the flux projector 
\begin{align}
    \ket{J(\vec{n})} = \sum_{l=1}^L \sum_{m=1}^{M_l} r_l \sigma_{lm} \{ &\Omega(\vec{n}, \vec{q}_{lm}, \vec{p}_{lm}) \ket{\vec{n} + \vec{q}_{lm} - \vec{p}_{lm}} \nonumber \\  
    - &\Omega(\vec{n}, \vec{q}_{lm}, \vec{q}_{lm}) \ket{\vec{n}} \}.
\label{eq:rulesME}
\end{align}

\subsection{Macroscopic master equation for the homopolymer}

\begin{figure*}[t]
    \centering
    \fbox{\includegraphics[width=0.97\textwidth]{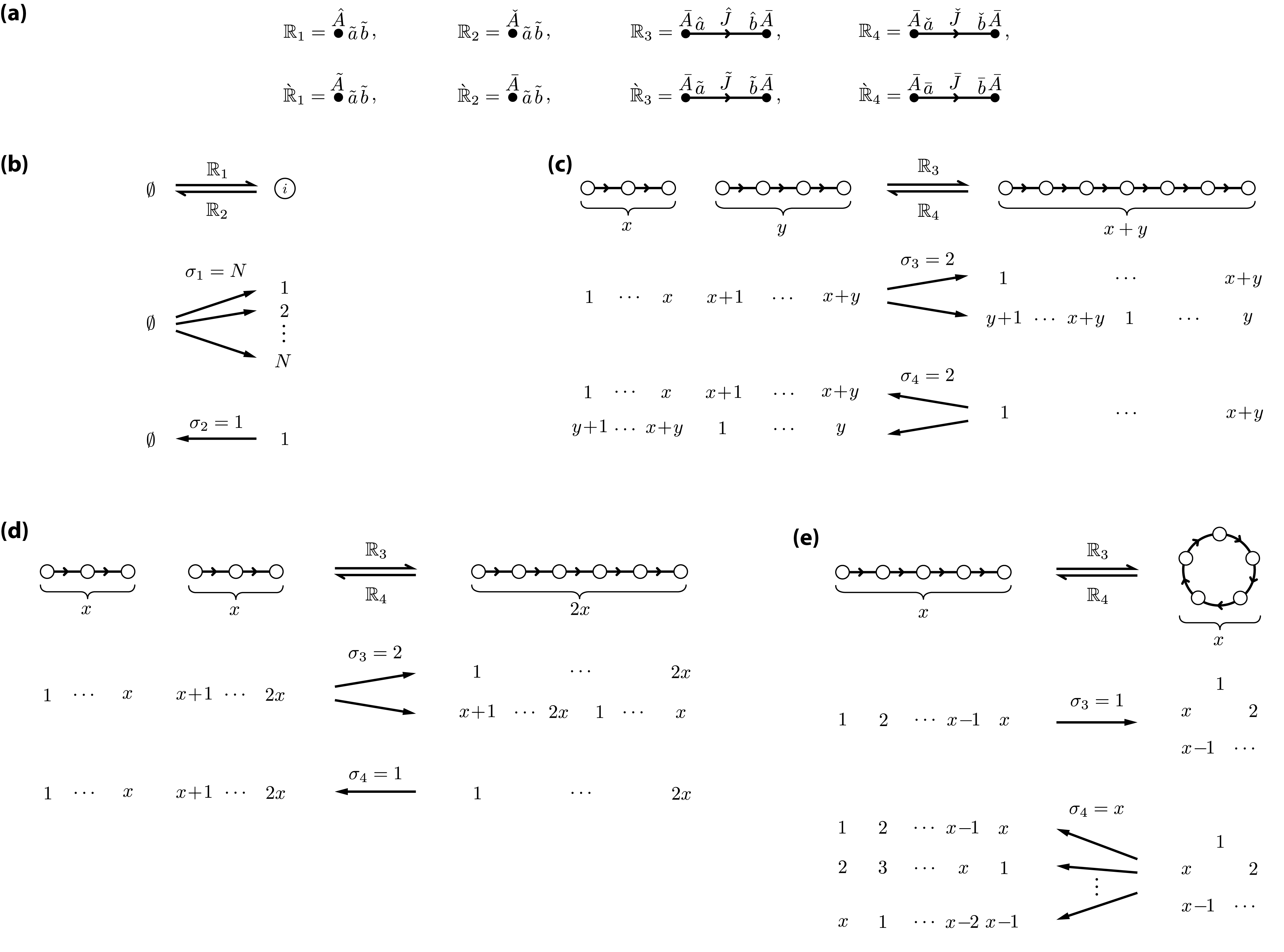}}
    \caption{The eight types of species-specific reactions that occur in the nonequilibrium homopolymer system. (a) Diagrammatic representations of the four reaction rule operators and their corresponding depletion operators. (b) $\mathbb{R}_1$ maps the vacuum state to $N$ different monomer microstates, while $\mathbb{R}_2$ maps each monomer microstate to a single vacuum state. (c-e) The three types of species-specific reactions effected by $\mathbb{R}_3$ and $\mathbb{R}_4$. (c) $\mathbb{R}_3$ can link an $x$-chain and $y$-chain together in two different ways, while $\mathbb{R}_4$ can split an $(x+y)$-chain into an $x$-chain and $y$-chain in two different ways. (d) $\mathbb{R}_3$ can link two $x$-chains together in two different ways, while $\mathbb{R}_4$ can split a $2x$-chain into two $x$-chains in only one way.(e) $\mathbb{R}_3$ can circularize an $x$-chain in only one way, while $\mathbb{R}_4$ can linearize an $x$-ring in $x$ different ways. Nodes are shown as open dots in panels (b-e) to indicate specific internal states; the numbers below each node indicate example values for the internal index $i$ of each component particle. 
    }
    \label{fig:homopolymer_noneq}
\end{figure*}

We now use our rule-based formalism to derive the macrostate master equation for the homopolymer system. Out of equilibrium, the dynamics of this system can be defined by $L=4$ reaction rule operators:
\begin{align}
    \mathbb{R}_1 = \sum_i \hat{A}_i \tilde{a}_i \tilde{b}_i,~~~~\mathbb{R}_3 = \sum_{i,j}  \hat{J}_{ij} \hat{a}_i \hat{b}_j \bar{A}_i \bar{A}_j, \nonumber  \\
    \mathbb{R}_2 = \sum_i \check{A}_i \tilde{a}_i \tilde{b}_i,~~~~\mathbb{R}_4 = \sum_{i,j}  \check{J}_{ij} \check{a}_i \check{b}_j \bar{A}_i \bar{A}_j.
\end{align}
Here, $\mathbb{R}_1$ creates a monomeric particle, $\mathbb{R}_2 = \mathbb{R}_1^\dagger$ destroys a monomeric particle, $\mathbb{R}_3$ creates an interaction between two particles, and $\mathbb{R}_4 = \mathbb{R}_3^\dagger$ destroys an interaction. Note that these four operators are given by the factory operators of Section \ref{sec:homopol} and their conjugates. The corresponding depletion operators are
\begin{align}
    \grave{\mathbb{R}}_1 = \sum_i \tilde{A}_i \tilde{a}_i \tilde{b}_i,~~~~ \grave{\mathbb{R}}_3 = \sum_{i,j}  \tilde{J}_{ij} \tilde{a}_i \tilde{b}_j \bar{A}_i \bar{A}_j, \nonumber \\
     \grave{\mathbb{R}}_2 = \sum_i \bar{A}_i \tilde{a}_i \tilde{b}_i,~~~~ \grave{\mathbb{R}}_4 = \sum_{i,j}  \bar{J}_{ij} \bar{a}_i \bar{b}_j \bar{A}_i \bar{A}_j.
\end{align}
Diagrammatic representations of these rule operators and depletion operators are shown in \fig{fig:homopolymer_noneq}{(a)}.

In Section \ref{sec:homopol} we showed that the macrostates of the homopolymer system are given by
\begin{equation}
	\ket{\vec{c}, \vec{r}} = \ket{c_1, r_1, c_2, r_2, \ldots}  = \prod_{x=1}^\infty \frac{\hat{C}_x^{c_x}}{c_x!}\frac{\hat{R}_x^{r_x}}{r_x!} \ket{0},
\end{equation}
where $c_x$ and $r_x$ respectively indicate the number of chains and rings of length $x$. To compute the macroscopic master equation, we compute the flux projector
\begin{align}
	\ket{J(\vec{c}, \vec{r})} = \sum_{l=1}^4 r_l (\mathbb{R}^\dagger_l - \grave{\mathbb{R}}_l) \ket{\vec{c},\vec{r}}. \label{eq:J_for_homopolymer}
\end{align}

We now evaluate \eq{eq:J_for_homopolymer} term-by-term. To ease notation, we show in the macrostate only the elements of $\vec{c}$ and $\vec{r}$ that change upon application of each operator and denote the unchanged elements by ``$\ldots$''.  The $l=1$ term corresponds to monomer creation. As illustrated in \fig{fig:homopolymer_noneq}{(b)},  $\mathbb{R}_1$ maps a single microstate (corresponding to no reactants) to $\sigma_1 \approx N$ microstates (corresponding to all possible monomer states). By \eq{eq:R_on_n} and \eq{eq:Rdep_on_n}, 
\begin{align}
	(\mathbb{R}^\dagger_1 -  \grave{\mathbb{R}}_1) \ket{\vec{c}, \vec{r}}\approx N \{ \ket{c_1\!-\!1,\ldots} -  \ket{\ldots} \}.
\end{align}
$\mathbb{R}_2$ maps a single monomeric particle microstate to $\sigma_2 = 1$ microstate (i.e., no products), and so
\begin{align}
    (\mathbb{R}^\dagger_2 - \grave{\mathbb{R}}_2) \ket{\vec{c}, \vec{r}} = (c_1\!+\!1) \ket{c_1\!+\!1,\ldots} - c_1 \ket{\ldots}.
\end{align}
The effects of $\mathbb{R}_3$ and $\mathbb{R}_4$ are more complex. $\mathbb{R}_3$ can effect three different types of reactions depending on the reactants. First, $\mathbb{R}_3$ can join together an $x$-chain and $y$-chain ($x < y$) to get an $(x+y)$-chain; this can be done in $\sigma_{3} = 2$ different ways [\fig{fig:homopolymer_noneq}{(c)}]. Second, $\mathbb{R}_3$ can can join together two $x$-chains to get a $2x$-chain; this can be done in $\sigma_{3} = 2$ ways [\fig{fig:homopolymer_noneq}{(d)}]. Third, $\mathbb{R}_3$ can join together the ends of an $x$-chain to get an $x$-ring; this can be done in only $\sigma_{3} = 1$ way [\fig{fig:homopolymer_noneq}{(e)}]. We therefore find that
\begin{widetext}
\begin{align}
    (\mathbb{R}_3^\dagger - \grave{\mathbb{R}}_3) \ket{\vec{c}, \vec{r}}  
        &= \sum_{x<y} 2 \{ (c_x\!+\!1)(c_y\!+\!1) \ket{c_x\!+\!1, c_y\!+\!1, c_{x+y}\!-\!1, \ldots} - c_x c_y \ket{\ldots} \} \nonumber \\
	&~~~+ \sum_x 2 \left\{ \frac{ (c_x\!+\!2)(c_x\!+\!1)}{2} \ket{c_x\!+\!2, c_{2x}\!-\!1, \ldots}  - \frac{c_x(c_x\!-\!1)}{2}\ket{\ldots} \right\} \nonumber \\
        &~~~+ \sum_x \{ (c_x\!+\!1) \ket{c_x\!+\!1, r_x\!-\!1, \ldots} - c_x \ket{\ldots} \}. 
\end{align}
Similarly, the inverse operator $\mathbb{R}_4$ can separate an $(x+y)$-chain into an $x$-chain and $y$-chain ($x < y$) in $\sigma_4 = 2$ ways, can separate a $2x$-chain into two $x$-chains in only $\sigma_4 = 1$ way, and can cut an $x$-ring to get an $x$-chain in $\sigma_4 = x$ different ways. Consequently,
\begin{align}
    (\mathbb{R}_4^\dagger - \grave{\mathbb{R}}_4) \ket{\vec{c}, \vec{r}} 
        &= \sum_{x<y} 2 \{ (c_{x+y}\!+\!1) \ket{c_x\!-\!1, c_y\!-\!1, c_{x+y}\!+\!1, \ldots}  - c_{x+y}\ket{\ldots} \} \nonumber \\
	& ~~~+ \sum_x \{ (c_{2x}\!+\!1) \ket{c_x\!-\!2, c_{2x}\!+\!1, \ldots} - c_{2x} \ket{\ldots} \} \nonumber \\
        & ~~~+ \sum_x x \{ (r_x\!+\!1) \ket{c_x\!-\!1, r_x\!+\!1, \ldots}. - r_x \ket{\ldots} \}.
\end{align}
The depletion terms in these expressions can be simplified as follows
\begin{align}
    \sum_{x<y} 2c_xc_y \!+\! \sum_x c_x(c_x\!-\!1) \!+\! \sum_x c_x  = n_\textrm{chain}^2,~~~~
   \sum_{x<y} 2c_{x+y} \!+\! \sum_x c_{2x} \!+\! \sum_x x r_x = n_\textrm{link},
\end{align}
where $n_\textrm{chain} = \sum_x c_x$ is the total number of chains and $n_\textrm{link} = \sum_x [(x-1)c_x + x r_x]$ is the number of links among all chains and rings. 

Evaluating the inner product $\braket{J(\vec{n}) | \psi(t)}$, we thus obtain the macroscopic master equation for the homopolymer:
\begin{align}
	\frac{d}{dt}P(\ldots) &= N r_1 P(c_1\!-\!1,\ldots) + r_2 (c_1\!+\!1) P(c_1\!+\!1,\ldots) \nonumber \\
	&+ r_3 \left[ \sum_{x<y} 2(c_x\!+\!1)(c_y\!+\!1) P(c_x\!+\!1,c_y\!+\!1,c_{x+y}\!-\!1,\ldots)  \right. \nonumber \\
	&~~~~~ \left. + \sum_x (c_x\!+\!2)(c_x\!+\!1)P(c_x\!+\!2,c_{2x}\!-\!1,\ldots) + \sum_x (c_x\!+\!1) P(c_x\!+\!1,r_x\!-\!1,\ldots) \right] \nonumber \\
	&+ r_4 \left[ \sum_{x<y} 2 (c_{x+y}\!+\!1) P(c_x\!-\!1, c_y\!-\!1, c_{x+y}\!+\!1,\ldots) + \sum_x (c_{2x}\!+\!1)P(c_x\!-\!2, c_{2x}\!+\!1,\ldots) \right. \nonumber \\
	&~~~~~ \left. + \sum_x x(r_x\!+\!1) P(c_x\!-\!1,r_x\!+\!1,\ldots) \right] \nonumber \\
	& - \left[ N r_1 + r_2 c_1 + r_3 n_\textrm{chain}^2 + r_4 n_\textrm{link} \right] P(\ldots).
\end{align}
\end{widetext}
To verify this result we focus on the depletion term, i.e., the coefficient of $P(\vec{c},\vec{r})$. The overall monomer creation rate is $N r_1$. This makes sense, as $r_1$ is the per-mode rate of excitation of the $A$ field. $r_1$ must therefore scale as $V/N$. The $r_2$ term reflects our assumption that only $A$ particles engaging in no interactions are able to be annihilated; it does not scale with $V$ or $N$. The $r_3$ term reflects the fact that new interactions form between the $a$-end of one chain and the $b$-end of either another chain or the same chain ($r_3$ scales as $V^{-1}$). The total number of available reactants in the system is therefore $n_\textrm{chain}^2$. The $r_4$ term reflects the assumption that any link can be annihilated regardless of the complex in which it occurs ($r_4$ does not scale with $V$ or $N$).

\section{\label{sec:simulations}Simulations}
\begin{figure*}[t]
    \centering    \includegraphics[width=2\columnwidth]{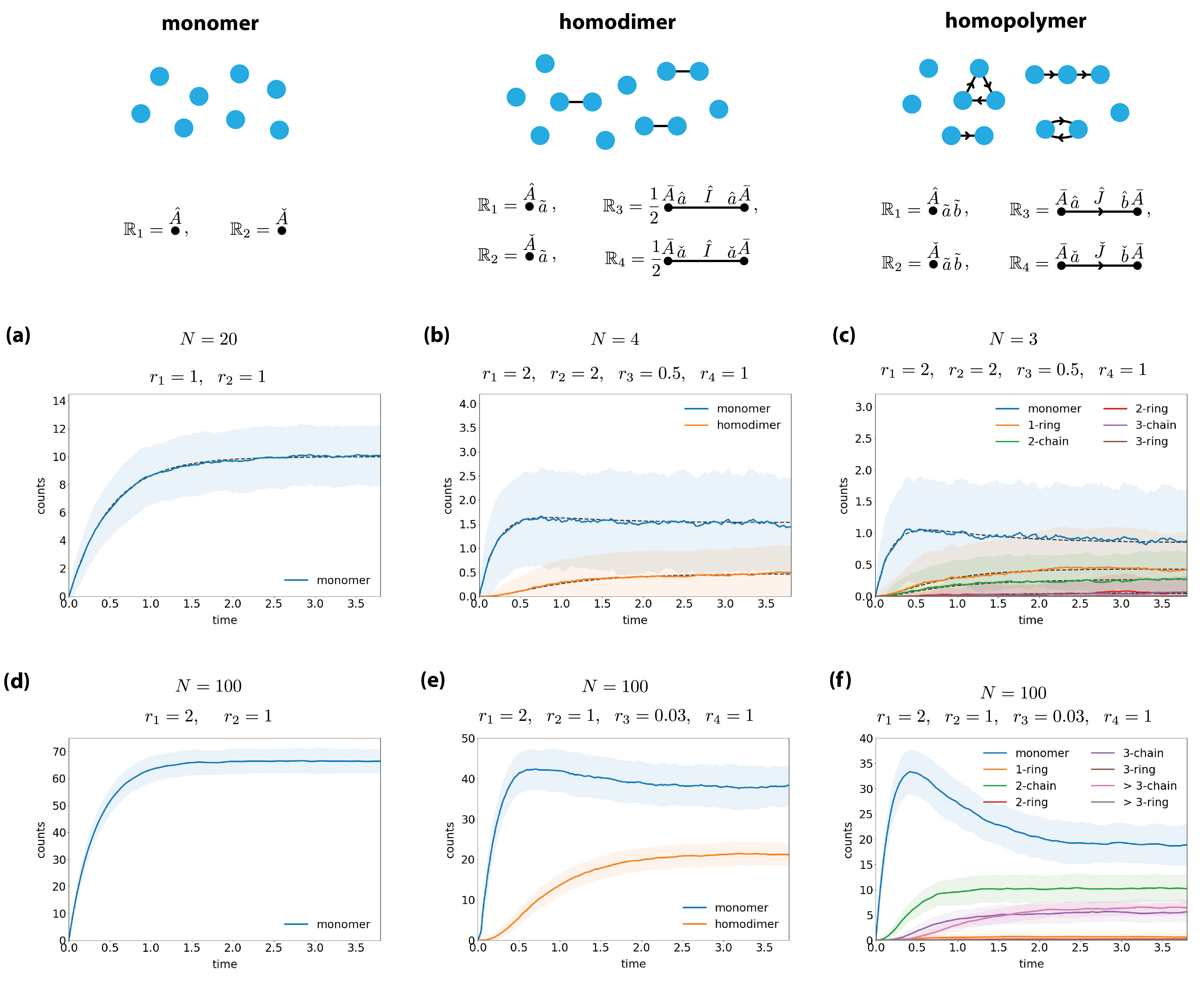}
    \caption{Simulations of nonequilibrium systems. (a-c) Deterministic and stochastic simulations for (a) a monomer system ($N=20$), (b) a homodimer system ($N=4$), and (c) a homopolymer system ($N=3$). (d-f) Stochastic simulations for the same systems as in panels (a-c) using $N=100$. Black dashed lines indicate mean abundances from deterministic simulations.  Solid lines plot mean abundances from 500 stochastic simulations. Error bands indicate standard deviations in abundance across the stochastic simulations.}
    \label{fig:simulations}
\end{figure*}

\subsection{Deterministic simulations}

Our formalism enables the numerical solution of the microstate master equation -- at least when $N$ is sufficiently small. One first defines a set of modes $\calM$, as well as a set of mode-specific rules $\mathcal{R} = \set{(\bbR_{li},r_{l})}$, where $l$ indexes the qualitatively different rules as in \eq{eq:W_from_R},  and $i$ indexes the internal states of the particles that each rule acts upon (so that $\bbR_l = \sum_i \bbR_{li}$). The transition operator is then computed using
\begin{equation}
    \mathbb{W} = \sum_{l=1}^L r_l \sum_i \left( \bbR_{li} - \grave{\bbR}_{li} \right).
\end{equation}
Given an initial state vector $\ket{\psi(0)}$, the state at time $t$ is computed using
\begin{equation}
    \ket{\psi(t)} = \exp \left\{ t\mathbb{W} \right\} \ket{\psi(0)}. \label{eq:integrated_master_equation}
\end{equation}
\figpanels{fig:simulations}{(a)}{(c)} show the results of this computation for three example systems: a monomer system ($N=20$), a homodimer system ($N=4$), and a homopolymer system ($N=3$). 

The primary limitation of this approach is that it requires computations involving very large vectors and matrices: $\ket{\psi}$ is a $2^{|\calM|}$-dimensional vector,  $\mathbb{W}$ is a $2^{|\calM|} \times 2^{|\calM|}$ matrix, and $|\calM|$ is polynomial in $N$, e.g., $|\calM| = N$ for the monomer, $|\calM| = 2N + {N \choose 2}$ for the homodimer, and $|\calM| = 3N + N^2$ for the homopolymer. Even using sparse matrix methods, we have found the direct evaluation of \eq{eq:integrated_master_equation} to be impractical for all but very small values of $N$. Nevertheless, these deterministic simulations provide a valuable way to check the accuracy of the stochastic simulations that we now turn to.

\subsection{Stochastic simulations}\label{subsec:sim}

Our formalism also enables stochastic simulations using the Gillespie algorithm \cite{Gillespie:1976aa,Gillespie:1977aa}. Importantly, these stochastic simulations can be carried out using much larger values of $N$. Algorithm \ref{alg:one} is one algorithm that does this. After explaining how the algorithm works, we illustrate its operator stepping through one iteration of the algorithm for a homodimer system. We then present computational results obtained using this algorithm and discuss the algorithm's current limitations. 

The input to Algorithm \ref{alg:one} consists of two strings, $s_\mytrans$ and $s_\myinit$. The string $s_\mytrans$ specifies the set of rules and their corresponding rates, while $s_\myinit$ specifies the initial state of the system, $\ket{s_0}$. The output of the algorithm is a trajectory object $\calT$, the downstream parsing of which provides time traces for the abundances of all single particles and complexes. Post-hoc processing of $\calT$ then provides time traces for all species of complex. We emphasize that this algorithm only tracks the excitation states of modes. In particular, there is no need during the execution of the algorithm to enumerate the different possible species of complex or even to track which species occur. Rather, time traces for the abundances of different complexes are determined only during the post-processing of $\calT$.

After processing its inputs, the algorithm sets time $t$ to zero and the trajectory object $\calT$ to be the empty set. Next, the function \FInitialize takes $s_\mytrans$ and $s_\myinit$ as inputs and outputs four sets of objects: 
\begin{itemize}
    \item $\calO$ is the set of all mode-specific operators (i.e., creation, annihilation, presence, and absence operators) for all fields. Every  $\bbO \in \calO$ has the attributes $\bbO.mode$, $\bbO.rules$, and $\bbO.eligible$. $\bbO.mode$ is a reference to the operator's mode $M \in \calM$. $\bbO.rules$ is a set of references to the rules $\bbR \in \calR$ that include $\bbO$ in their operator product. $\bbO.eligible$ is a Boolean flag indicating whether $\bbO$ can be applied to $\ket{s_t}$ without killing it. 
    \item $\calM$ is the set of all modes for all fields. Every $M \in \calM$ has the attributes $M.operators$ and $M.excited$. $M.operators$ is a set of references to the four mode-specific operators (i.e., $\hat{M}$, $\check{M}$, $\bar{M}$, and $\tilde{M}$). $M.excited$ is a Boolean value indicating whether the mode is excited. If TRUE, $\hat{M}$ and $\tilde{M}$ are ineligible, while  $\check{M}$ and $\bar{M}$ are eligible. If FALSE, $\hat{M}$ and $\tilde{M}$ are eligible, whereas  $\check{M}$ and $\bar{M}$ are ineligible.
    \item $\calR$ is the set of all mode-specific rules, each defined as a product of mode-specific operators. Every rule $\bbR \in \calR$ has the attributes $\bbR.operators$ and $\bbR.rate$. $\bbR.operators$ is a set of references to the mode-specific operators $\bbO \in \calO$ that comprise $\bbR$. $\bbR.rate$ is the rate at which the rule is applied when eligible.
    \item $\calC$ is the ``state constructor,'' i.e., the set of creation operators that, when applied to $\ket{0}$, yield $\ket{s_t}$, the system state at time $t$. The specific $\calC$ returned by \FInitialize corresponds to $\ket{s_0}$, which is specified by the string $s_\myinit$. In particular, if the user specifies that $\ket{s_0} = \ket{0}$, then $\calC = \set{}$. 
\end{itemize}
$\calO$, $\calM$, and $\calR$ are static, whereas $\calC$ evolves in time. Also note that
\begin{eqnarray}
	M = \bbO.mode &\Leftrightarrow& \bbO \in M.operators, \\
	\bbR \in \bbO.rules &\Leftrightarrow& \bbO \in \bbR.operators
\end{eqnarray}
for all $\bbO \in \calO$, $M \in \calM$, and $\bbR \in \calR$.

%
%
\begin{algorithm}[t!]
\DontPrintSemicolon
\caption{Gillespie simulation}\label{alg:one}
\KwData{$s_\mytrans$, $s_\myinit$, $n_{\max}$}
\KwResult{$\calT$}

\BlankLine
\tcc{Initialize time, trajectory, operators, modes, rules, and state constructor.}
$t \gets 0$\; \label{algline:tau_0}
$\calT \gets \set{}$\;
$\calO, \calM, \calR, \calC \leftarrow \FInitialize(s_\mytrans, s_\myinit)$\; \label{algline:init}
\BlankLine
\tcc{Carry out Gillespie algorithm.}
\For{$n \in \set{1, \ldots, n_{\max}}$}{
	$\calR^* \gets \set{\bbR: \bbR \in \calR, \FIsRuleEligible{$\bbR$}}$\; \label{algline:eligible_rules}
	$\bbR, \Delta t \leftarrow \FGillespieStep(\calR^*)$\; \label{algline:sample_rule}
	$t \gets t + \Delta t$\;
	\For{$\bbO \in \bbR.operators$}{\label{algline:Cupdatestart}
		\If{$\bbO.type == \textrm{``creation''}$}{
			$\calC \leftarrow \calC \cup \set{\bbO}$\;
			$\FFlipModeExcitation(\bbO.mode)$\;
		}
		\If{$\bbO.type == \textrm{``annihilation''}$}{
			$\calC \leftarrow \calC \setminus \set{\bbO^\dagger}$\;
			$\FFlipModeExcitation(\bbO.mode) \label{algline:Cupdateend} $\;
		}
	}
	$\calT \leftarrow \calT \cup (n, t, \bbR, \calC)$\;
}

\BlankLine
\tcc{Create operators, modes, rules, and initial state constructor.}
\Fn{\FInitialize{$s_\mytrans$, $s_\myinit$}}{
        \ldots\;
	\KwRet{$\calO$, $\calM$, $\calR$, $\calC$}\;
}

\BlankLine
\tcc{Compute whether rule is eligible.}
\Fn{\FIsRuleEligible{$\bbR$}}{ \label{algline:isruleeleigible}
    \KwRet{\FProd{$\set{\bbO.eligible: \bbO \in \bbR.operators}$}}\;
}

\BlankLine
\tcc{Randomly choose rule and time step using the Gillespie algorithm.}
\Fn{\FGillespieStep{$\calR^*$}}{ \label{algline:gillespie_step}
	$\vec{r} \gets \set{\bbR.rate: \bbR \in \calR^*}$\;
	$r_{\rm tot} \gets \FSum(\vec{r})$\;
	$\Delta t \gets $\,\FExponential{$rate=r_{\rm tot}$}\;
	$\bbR \gets $\,\FChoose{$set=\calR^*$, $weights=\vec{r}$}\;
	\KwRet{$\bbR, \Delta t$}\;
}

\BlankLine
\tcc{Flip excitation state of mode and eligibility of corresponding operators.}
\Fn{\FFlipModeExcitation{$M$}}{ 
	$M.excited \gets \mathrm{NOT}~M.excited$\;
	\For{$\bbO \in M.operators$}{
		$\bbO.eligible \gets \mathrm{NOT}~\bbO.eligible$\;
	}
}
\end{algorithm}

After initialization, Algorithm \ref{alg:one} enters a while loop. Each execution of the loop applies one rule $\bbR$ to the system state $\ket{s_t}$, then steps the system forward in time from $t$ to $t + \Delta t$. The contents of this while loop are as follows:
\begin{itemize}
	\item Line \ref{algline:eligible_rules} identifies the set of  eligible rules and saves them in set $\calR^*$. This is the most computationally expensive part of Algorithm \ref{alg:one}, since it must loop through every possible rule and test that rule for eligibility using the function \FIsRuleEligible. A rule $\bbR$ is eligible if and only if every operator in $\bbR.operators$ is eligible.  
	\item Line \ref{algline:sample_rule} uses the rates of the rules in $\calR^*$ to randomly sample a time step $\Delta t$ and a corresponding rule $\bbR$ via the Gillespie algorithm, which is implemented by \FGillespieStep. Time $t$ is then incremented by $\Delta t$.
	\item Lines \ref{algline:Cupdatestart}-\ref{algline:Cupdateend} update the state constructor $\calC$ and the excitation state of modes affected by the rule $\bbR$. To do this, a for loop is carried out over all operators $\bbO \in \bbR.operators$. If $\bbO$ is a creation operator, then $\bbO$ is added to the state constructor $\calC$. Alternatively, if $\bbO$ is an annihilation operator, then the corresponding creation operator  $\bbO^\dagger$ is removed from the state constructor $\calC$. In either case, \FFlipModeExcitation is applied to the mode $M=\bbO.\textrm{mode}$. This flips the $excitation$ attribute of $M$, as well as the $eligible$ attribute of all operators in $M.operators$.  
	\item Finally, the loop stores a tuple reporting the updated time $t$, the rule $\bbR$, and the resulting state constructor $\calC$ in the trajectory object $\calT$.
\end{itemize}

We now illustrate how this algorithm works by following it through initialization and one execution of the while loop. Assume that the string $s_\mytrans$ specifies the  following kinetic rules for a homodimer system:
\begin{align}
    \bbR_1 &= \sum_i \hat{A}_i \tilde{a}_i,~~~~\bbR_3 = \sum_{i < j}  \bar{A}_i \bar{A}_j \hat{a}_i \hat{a}_j \hat{I}_{ij}, \nonumber  \\
    \bbR_2 &= \sum_i \check{A}_i \tilde{a}_i,~~~~\bbR_4 = \sum_{i < j}  \bar{A}_i \bar{A}_j  \check{a}_i \check{a}_j \check{I}_{ij}.
\end{align}
We further suppose that the string $s_\myinit$ specifies an initial state containing four monomers having indices 2, 3, 5, and 7, i.e., 
\begin{equation}
\ket{s_0} = \hat{A}_2 \hat{A}_3 \hat{A}_5 \hat{A}_7 \ket{0}.
\end{equation}
Taking $s_\mytrans$ and $s_\myinit$ as input, the function \FInitialize returns the following sets of objects:
\begin{eqnarray}
\calM &=& \set{A_i}_i \cup \set{a_i}_i \cup \set{I_{ij}}_{i < j}, \nonumber \\
\calO &=& \set{\hat{A}_i, \check{A}_i, \bar{A}_i, \tilde{A}_i}_i \cup 
      \set{\hat{a}_i, \check{a}_i, \bar{a}_i, \tilde{a}_i}_i \cup \nonumber \\
  & & \set{\hat{I}_{ij}, \check{I}_{ij}, \bar{I}_{ij}, \tilde{I}_{ij}}_{i < j}, \nonumber \\
\calR &=& \set{\hat{A}_i \tilde{a}_i}_i \cup \set{\check{A}_i \tilde{a}_i}_i \cup \nonumber \\
  & & \set{\bar{A}_i \bar{A}_j \hat{a}_i \hat{a}_j \hat{I}_{ij}}_{i<j} \cup 
   \set{\bar{A}_i \bar{A}_j \check{a}_i \check{a}_j \check{I}_{ij}}_{i<j}, \nonumber \\
\calC &=& \set{\hat{A}_2, \hat{A}_3, \hat{A}_5, \hat{A}_7}.
\end{eqnarray}
Here the indices $i$ and $j$ are understood to run over $1, \ldots, N$, and the rates $r_l$ corresponding to each $\bbR_{li}$ are kept implicit. \FInitialize also sets the values of $M.excited$ for all $M \in \calM$, and of $\bbO.eligible$ for all $\bbO \in \calO$. $M.excited$ is TRUE for modes $A_2$, $A_3$, $A_5$, and $A_7$, and is FALSE for all other modes (including all modes of the fields $a$ and $I$). Consequently, the set $\calO^*$ of eligible operators is 
\begin{eqnarray}
    \calO^* &=& \set{\hat{A}_i,   \tilde{A}_i}_{i \notin \set{2, 3, 5, 7}} \cup 
            \set{\check{A}_i, \bar{A}_i}_{i \in \set{2, 3, 5, 7}} \cup \nonumber \\
           & & \set{\hat{a}_i,   \tilde{a}_i}_i \cup 
            \set{\hat{I}_{ij}, \tilde{I}_{ij}}_{i < j}. 
\end{eqnarray}
Now consider the first execution of the while loop. In line \ref{algline:eligible_rules} of Algorithm 1, \FIsRuleEligible is evaluated on every rule $\bbR \in \calR$. Based on the eligibility of each operator in $\bbR.operators$, the eligible rules are found to be
\begin{eqnarray}
    \calR^* &=& \set{\hat{A}_i \tilde{a}_i}_{i \notin \set{2, 3, 5, 7}} \cup 
  \set{\check{A}_i \bar{a}_i}_{i \in \set{2, 3, 5, 7}} \cup \nonumber \\
  & & \set{\bar{A}_i \bar{A}_j \hat{a}_i \hat{a}_j \hat{I}_{ij}}_{i<j \in \set{2,3,5,7}}. \label{eq:first_calR}
\end{eqnarray}
Next, \FGillespieStep chooses a random eligible rule $\bbR \in \calR^*$ and a time increment $\Delta t$ that is then added to $t$. Suppose
\begin{equation}
\bbR = \bar{A}_3 \bar{A}_7 \hat{a}_3 \hat{a}_7 \hat{I}_{37}
\end{equation}
is chosen. Applying $\bbR$ to the initial state vector then yields the updated state vector,
\begin{equation}
\ket{s_{\Delta t}} = \bbR \ket{s_0} = \hat{A}_2 \hat{A}_3 \hat{A}_5 \hat{A}_7 \hat{a}_3 \hat{a}_7 \hat{I}_{37} \ket{0}.
\end{equation}
To register this change, the state constructor is updated to 
\begin{equation}
\calC = \set{\hat{A}_2, \hat{A}_3, \hat{A}_5, \hat{A}_7, \hat{a}_3, \hat{a}_7, \hat{I}_{37}}.
\end{equation}
Calls to \FFlipModeExcitation then set the $excited$ attributes of modes $a_3$, $a_7$, $I_{37}$ to TRUE and flip the $eligible$ attribute of the four operators corresponding to each of these modes. The resulting set of eligible operators is
\begin{eqnarray}
    \calO^* &=& \set{\hat{A}_i,  \tilde{A}_i}_{i \notin \set{2, 3, 5, 7}} \cup  \set{\check{A}_i, \bar{A}_i}_{i \in \set{2, 3, 5, 7}} \cup \nonumber \\
           & & \set{\hat{a}_i,   \tilde{a}_i}_{i \notin \set{3,7}} \cup \set{\check{a}_i, \bar{a}_i}_{i \in \set{3,7}} \cup \nonumber  \\
           & & \set{\hat{I}_{25}, \tilde{I}_{25}, \check{I}_{37}, \bar{I}_{37},}.
\end{eqnarray}
Finally the updated time $t$, the chosen rule $\bbR$, and the updated state constructor $\calC$ are added as a tuple to the trajectory $\calT$. 

In the next execution of the while loop, the set $\calR^*$ of eligible rules is found by \FIsRuleEligible to be
\begin{eqnarray}
    \calR^* &=& \set{\hat{A}_i \tilde{a}_i}_{i \notin \set{2, 3, 5, 7}} \cup  \set{\check{A}_i \tilde{a}_i}_{i \in \set{2, 5}} \cup \nonumber \\
  & & \set{\bar{A}_2 \bar{A}_5 \hat{a}_2 \hat{a}_5 \hat{I}_{25},\, \bar{A}_3 \bar{A}_7 \check{a}_3 \check{a}_7 \check{I}_{37}}. 
\end{eqnarray}
It is worth noting the changes to $\calR^*$ vs. \eq{eq:first_calR}. The monomer annihilation rules $\check{A}_3 \tilde{a}_3$ and $\check{A}_7 \tilde{a}_7$ have been removed because the modes $a_3$ and $a_7$ are now excited, making the operators $\tilde{a}_3$ and $\tilde{a}_7$ ineligible. This prevents monomers joined by interactions from being annihilated, thereby leaving ``dangling'' interactions. In addition, all interaction creation rules except $\bar{A}_2 \bar{A}_5 \hat{a}_2 \hat{a}_5 \hat{I}_{25}$ have become ineligible. This prevents the $A_3$ and $A_7$ monomers, which are already interacting with each other, from participating in multiple interactions. Finally, the interaction annihilation rule $\bar{A}_3 \bar{A}_7 \check{a}_3 \check{a}_7 \check{I}_{37}$ becomes eligible, allowing the newly-formed bond to dissociate. 

\fig{fig:simulations}{} shows this stochastic algorithm applied to monomer, homodimer, and homopolymer systems. Panels (a-c) validate this algorithm by showing that the mean abundance of each species found across 500 simulations closely traces the mean abundance predicted by the deterministic algorithm. As noted above, however, these comparisons can only be carried out at small $N$ due to limitations of the deterministic algorithm. This stochastic algorithm can be performed at much larger values of $N$, e.g., \figpanels{fig:simulations}{(d)}{(f)} show such simulations performed using $N = 100$. 

The size of $N$ is still a limitation for Algorithm \ref{alg:one}. The bottleneck is line 5, which requires iterating through all mode-specific rules in $\calR$ and testing each one for eligibility. This step takes $o(|\calR|)$ time, and $|\calR|$ is polynomial in $N$:  $|\calR| = 2N$ for the monomer, $|\calR| = 2N + 2{N \choose 2}$ for the homodimer, and  $|\calR| = 2N + 2N^2$ for the homopolymer. That said, Algorithm \ref{alg:one} was developed only as proof-of-principle and has not been optimized for efficiency. Indeed, we expect the bottleneck can be eliminated by using more sophisticated methods for tracking which operators and rules are eligible given the state constructor, thereby enabling simulations using arbitrarily large values of $N$.

\section{\label{sec:expressiveness}Expressiveness}

\newcommand{\Ztree}{Z_\textrm{tree}}
\newcommand{\Zxgrove}{Z_\textrm{$x$-grove}}
\newcommand{\Zchain}[1]{Z_\textrm{$#1$-chain}}
\newcommand{\Zring}[1]{Z_\textrm{$#1$-ring}}

\begin{figure*}[t]
    \centering
    \includegraphics[width=\textwidth]{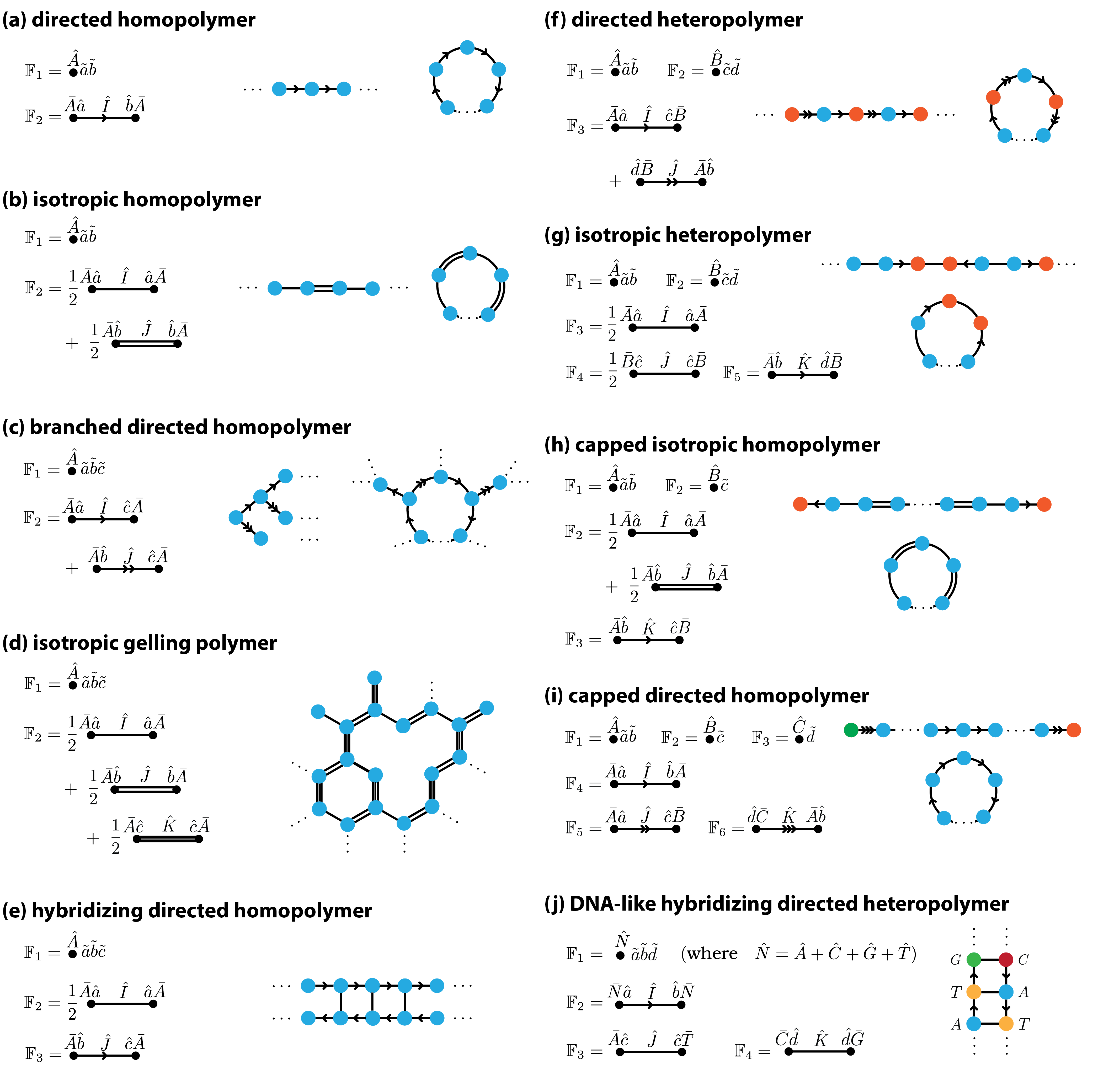}
    \caption{Factories defining various polymer systems. Each panel shows the factory operators used to define system in equilibrium.  The homopolymer system analyzed in Section \ref{sec:homopol} is shown in panel (a) for comparison.}
    \label{fig:various_polymers}
\end{figure*}

Our formalism provides a rule-based approach for defining and analyzing a diverse array of stochastic chemical systems in which multi-particle complexes can form. Here we illustrate the expressiveness of our formalism by briefly considering a variety of such systems, both in and out of equilibrium. 

In equilibrium, systems are defined by a factory $\vec{\mathbb{F}}$ and a Hamiltonian $\mathbb{H}$. Putting the Hamiltonian aside for the moment, it is interesting to consider the qualitatively different sets of complexes that can arise from simple changes to the factory. This expressiveness is perhaps most apparent in polymer systems. \fig{fig:various_polymers}{} shows nine such polymeric systems derived from variations on the homodimer and homopolymer systems described in previous sections. Two such systems, the isotropic homopolymer [\fig{fig:various_polymers}{(b)}] and branched homopolymer [\fig{fig:various_polymers}{(c)}] are further analyzed below. 

\begin{figure*}[p]
    \centering
    \includegraphics[width=\textwidth]{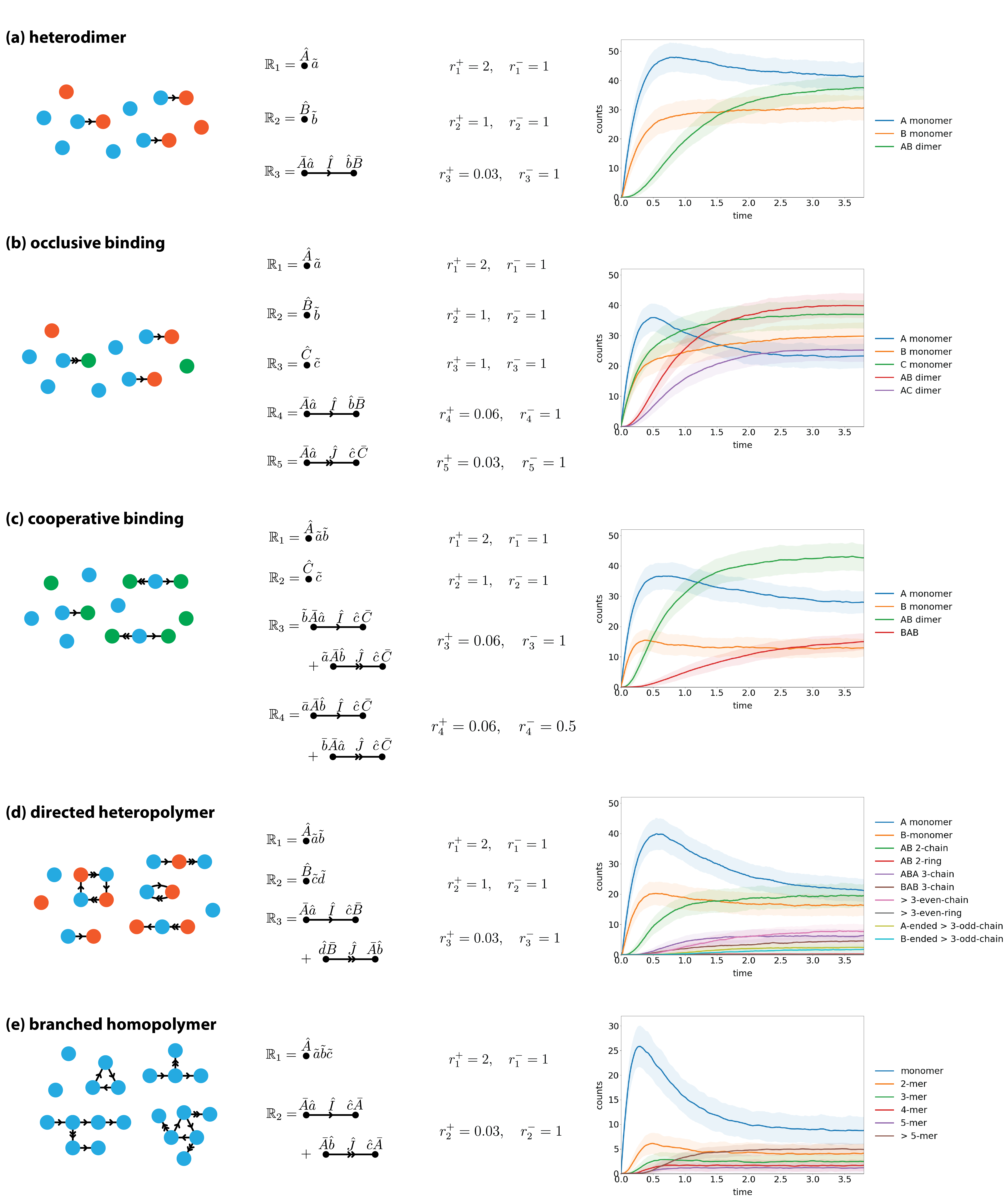}
    \caption{Stochastic simulations of various nonequilibrium models. Shown are results for (a) a heterodimer system, (b) an occlusive binding system, (c) a cooperative binding system, (d) a heteropolymer system, and (e) a branched homopolymer system. For each system, $r_l^+$ denotes the rate at which the forward rule, $\mathbb{R}_l$, is applied, while $r_l^-$ denotes the rate at which the reverse rule, $\mathbb{R}_l^\dagger$, is applied. Solid lines indicate mean abundances from 500 stochastic simulations using $N=100$. Error bands indicate abundance standard deviations across the simulations.}
    \label{fig:various_simulations}
\end{figure*}

Out of equilibrium, systems are defined by a set of rules, with each rule $\mathbb{R}_l$ assigned a corresponding rate $r_l$. \fig{fig:various_simulations}{} shows five such nonequilibrium systems. These rules again derive from variations on the homodimer and homopolymer systems. \fig{fig:various_simulations}{} also shows the results of stochastic simulations carried out using Algorithm \ref{alg:one} for specific choices of the rate parameters. 

\subsection{Isotropic homopolymer}

We now analyze the isotropic homopolymer shown in \fig{fig:various_polymers}{(b)}. This system is defined by one species of monomeric subunit ($A$) with two sites ($a$ and $b$) capable of participating in two classes of symmetric interaction ($I$ and $J$). We aim to compute the partition function for the system assuming the Hamiltonian
\begin{align}
    \mathbb{H} = - \mu \sum_i \bar{A} + \frac{\epsilon}{2} \sum_{i,j}(\bar{I}_{ij} + \bar{J}_{ij}).
\end{align}

As with the homopolymer of Section \ref{sec:homopol}, the two factory operators for this system generate chains and rings. However, the use of two distinct types of symmetric interaction complicates these species. First, there are no self-interacting monomers. Second, the $I$ and $J$ interactions alternate in all multimers. For $x$-chains this means that, when $x$ is even, there are two different species related by the exchange of $I$ and $J$ interactions. When $x$ is odd, there is instead only one-species of $x$-chain, as exchanging $I$ and $J$ is equivalent to flipping the order of the subunit indices. Moreover, $x$-rings occur only when $x$ is even. These rings have $x/2$-fold rotational symmetry, and for $x \geq$ 4 they additionally have 2-fold mirror symmetry. 

The log partition function density of the system is therefore given by 
\begin{align}
    \frac{\log Z}{V} &= \sum_{x=1}^\infty \log \Zchain{(2x\!-\!1)} + \sum_{x=1}^\infty \log \Zchain{2x} \nonumber \\
    & ~~~~ + \log \Zring{2} + \sum_{x=2}^\infty \log \Zring{2x}. \label{eq:Z_isotorpichomopolymer}
\end{align}
In terms of the effective chemical potential $\mu' = \mu + k_B T \log \frac{N}{V}$ and effective interaction energy $\epsilon' = \epsilon - k_B T \log V$, as well as the control parameter $\eta = e^{\beta(\mu' - \epsilon')}$ (which represents the Boltzmann weight of a single particle with a dangling bond), the single-complex partition functions in \eq{eq:Z_isotorpichomopolymer} are 
\begin{align}
	\log\Zchain{(2x-1)} &= e^{\beta \epsilon'} \eta^{2x-1},~~~
	\log\Zchain{2x} = 2 e^{\beta \epsilon'} \eta^{2x}, \nonumber \\
	\log\Zring{2} &= \frac{\eta^{2}}{2V},~~~
	~~~~~~~~~\log\Zring{2x} = \frac{\eta^{2x}}{4xV}.
\end{align}
Summing the terms in \eq{eq:Z_isotorpichomopolymer} we find that, for $0 < \eta < 1$,
\begin{align}
	\frac{\log Z}{V} = \frac{e^{\beta \epsilon'}\eta}{1 - \eta} + \frac{e^{\beta \epsilon'}\eta^2}{1 - \eta^2} + \frac{\eta^{2}}{4V} - \frac{\log(1 - \eta^2)}{4V}.
\end{align} 
As with the homopolymer, the partition function diverges as $\eta \to 1$ from below. Defining $\delta = 1 - \eta$, we find that the concentration of $A$ particles diverges in this limit, scaling as
\begin{align}
    \frac{\braket{\bar{A}}}{V} \approx \frac{3 e^{\beta \epsilon'}}{2 \delta^2}. \label{eq:heteropolymer_scaling}
\end{align}
Again, this divergence is dominated by the $x$-chains even at finite $V$. The factor of $\frac{3}{2}$ difference between this result and the homopolymer result in \eq{eq:homopolymer_A_cond_div} reflects the fact that, while both even and odd chains contribute to \eq{eq:heteropolymer_scaling}, there are twice as many species of even chains (but the same number of odd chains) as in the homopolymer system.

\subsection{Branched directed homopolymer}

Consider now the branched directed homopolymer system shown in \fig{fig:various_polymers}{(c)}. This system is defined by one species of monomeric subunit ($A$) with three sites ($a$, $b$, and $c$) capable of participating in two classes of directed interaction ($I$ and $J$).  We assume that the system is in thermal equilibrium and is governed by the Hamiltonian
\begin{align}
    \mathbb{H} = - \mu \sum_i \bar{A} + \epsilon \sum_{i,j}(\bar{I}_{ij} + \bar{J}_{ij}).
\end{align}
The two factory operators for this system generate two broad classes of complex: ``trees,'' which branch out from a single $A$ in which site $c$ is unoccupied, and ``groves,'' which consist of multiple trees branching out from a central closed ring. The log partition function density of this system is therefore given by
\begin{align}
    \frac{\log Z}{V} &= \Ztree + \sum_{x=1}^\infty \Zxgrove, 
\end{align}
where $\Ztree$ is the partition function for all trees and $\Zxgrove$ is the partition function for all groves with trees extending off a central $x$-ring. To proceed, we let $\xi$ represent the partition function for a tree with a dangling interaction extending off the root. In terms of this quantity,
\begin{align}
     \Ztree = e^{\beta \epsilon'} \xi,~~~ \Zxgrove = \frac{2^x \eta^x \left( 1 + \xi \right)^x}{Vx}, 
\end{align}
where $\eta = e^{\beta(\mu'-\epsilon')}$ is again the Boltzmann weight of a single particle with a dangling bond. In $\Ztree$, the factor of $e^{\beta \epsilon'}$ removes the effect of the dangling bond. In $\Zxgrove$, the factor of $\eta^x$ accounts for the particles and bonds in each $x$-ring, the factor of $2^x$ accounts for the fact that each bond in the $x$-ring can be of type $I$ or $J$, and the $(1+\xi)^x$ factor accounts for the fact that each $A$ within the ring can either be bare or have a tree attached. As with the homopolymer, the factor of $\frac{1}{x}$ compensates for the rotational symmetry of the ring while $\frac{1}{V}$ reflects the entropic cost of self-circularization. 

We solve for $\xi$ by noting that the self-similar structure of each tree complex yields the recursion relation
\begin{align}
    \xi = \eta (1 + 2 \xi + \xi^2).
\end{align}
Solving this quadratic equation and using the limiting behavior $\xi \approx \eta$ as $\eta \to 0$, we derive
\begin{align}
    \xi = \frac{1 - 2 \eta - \sqrt{1 - 4 \eta}}{2 \eta}.
\end{align}
Expressing $\Ztree$ and $\Zxgrove$ in terms of $\xi$ and summing $\Zxgrove$ over all $x$, we find that for $0 < \eta < \frac{1}{4}$,
\begin{align}
    \frac{\log Z}{V} = e^{\beta \epsilon'} \left( \frac{1 - 2 \eta - \sqrt{1 - 4 \eta}}{2 \eta}\right) - \frac{\log(1 - 4 \eta)}{2V}.
\end{align}
As with the homopolymer system, the contribution from the circularized (i.e., $x$-grove) species vanishes in the $V \to \infty$ limit. We also find that the partition function is not defined for $\eta > \frac{1}{4}$. In particular, the mean concentration of $A$ particles is found to diverge as $\delta = \frac{1}{4} - \eta \to 0$ according to
\begin{align}
    \frac{\braket{\bar{A}}}{V} \approx \frac{e^{\beta \epsilon'}}{2 \sqrt{\delta}} + \frac{1}{8 V \delta},
\end{align}
with the first and second terms respectively corresponding to trees and groves. This result is qualitatively different from the corresponding results for the directed homopolymer in Section \ref{sec:homopol} and for the isotropic homopolymer analyzed above. In particular, the circularized species (the groves) dominate over the linear species (the trees) in the $\delta \to 0$ limit when $V$ is kept finite. The asymptotic behavior of the system therefore depends on which limit one takes first, $\eta \to \frac{1}{4}$ or $V \to \infty$. Moreover, the divergence is more mild than in the non-branched systems, scaling as either  $\delta^{-1/2}$ or $\delta^{-1}$ (depending on how one handles $V$) rather than $\delta^{-2}$. 

\section{\label{sec:discuss}Discussion}
We have introduced an algebraic formalism for the rule-based modeling of multi-particle complexes in stochastic chemical systems. This algebra is based on a Fock space that allows not only the creation and annihilation of particles, but also the joining of particles into complexes based on specified rules. The Fock space comprises three types of hard-core boson fields; these represent particles, particle-particle interactions, and occupied binding sites. We have also described a formal diagrammatic approach that facilitates the use of this algebra. 

For equilibrium systems, we showed that the set of all possible complexes can be rigorously specified by a ``factory'' and a Hamiltonian. The factory is an ordered set of operators that define how to construct complexes; the Hamiltonian is an operator that specifies rules for computing the Gibbs free energy of a complex based on its components. We showed for multiple systems how these rule-based definitions can be used to compute generating functions and partition functions, and to analyze scaling behavior near critical polymerization concentrations. 

For nonequilibrium systems, we showed how to rigorously specify system dynamics in a rule-based manner. Specifically, we showed how a set of reaction rule operators and corresponding rates can be used to define the transition matrix of a ``microstate'' master equation. From this transition matrix one can then analytically compute the corresponding ``macrostate'' master equation, which governs the time evolution of observables. We also developed a Gillespie algorithm for simulating stochastic chemical systems based on these rule operators and corresponding rates. 

The essential feature of our formalism, one that distinguishes it from previous approaches for modeling many-body systems of classical particles, is that it explicitly represents internal particle states. These internal states endow each particle with its own identity, thus allowing preexisting particles to join together into multi-particle complexes. Notably, our approach to modeling these internal states is consistent with the behavior of quantum systems in the decoherence limit. In this limit, the reduced density matrix for each particle becomes diagonal with elements along the diagonal quantifying the probability of each energy eigenstate \cite{Zurek:2003aa, Scholsshauer:2019aa}. The orthonormal microstates in our formalism correspond to these diagonal positions in the reduced density matrix (i.e., the energy eigenstates), and the probabilities that multiply these microstates correspond to the values of the reduced density matrix at these positions. Classical particles, such as proteins, have many distinct energy eigenstates corresponding to different internal excitation modes, and the internal states of particles in our formalism stand in for these modes. Our formalism assumes a specific number $N$ of such modes, but this choice does not affect results when $N$ is sufficiently large. 

We envision a variety of potential analytic applications for our formalism. As in the work of Doi \cite{Doi:1976a, Doi:1976b}, it may be possible to use the algebra we have introduced to carry out diagrammatic perturbation theory calculations. As in the work of Peliti \cite{Peliti:1985kp} and Goldenfeld \cite{Goldenfeld:1984}, it may also be possible to identify a path integral formulation of this algebra. While our analysis focused on zero-dimensional (i.e., well-mixed) systems, we expect that it should be straightforward to apply our formalism to spatially extended systems in which diffusion plays an important role (as in \cite{Doi:1976a, Doi:1976b, Peliti:1985kp, Goldenfeld:1984}). 

We also envision a variety of computational applications. Serious computational applications will require reworking our proof-of-principle algorithm (Algorithm \ref{alg:one}) so that its speed does not scale with $N$, but we expect this will be straightforward using more advanced bookkeeping methods. The resulting algorithm may provide advantages over existing rule-based modeling approaches, since the underlying objects in our formalism (hard core bosons) are simpler than those of existing algorithms (e.g., port graphs). Our formalism might also facilitate the development of qualitatively different computational strategies for rule-based modeling, e.g., the use of finite state projections \cite{Munsky:2006aa} or tensor networks \cite{Nicholson:2023aa} to approximate the solution of the master equation. 

In this paper we have focused on polymer systems, which best illustrate how large complexes can arise from simple interaction rules. Our initial motivation in pursuing this project, however, was in biological systems. We specifically sought to develop methods for modeling the biophysical mechanisms of gene regulation. Gene regulation is controlled by large protein-nucleic acid complexes \cite{Bintu:2005bn, Bintu:2005ur}, with individual regulatory sequences able to nucleate the formation of large numbers of different macromolecular assemblies. A major goal in this field is to understand these complexes as well as their effects on gene expression using biophysical models, and substantial progress has been made using both equilibrium models \cite{Shea:1985tz, Dodd:2004ex, Kuhlman:2007ts, Cui:2013ug} and nonequilibrium models \cite{Estrada:2016ct, Scholes:2017gz, Gunawardena:2020aa}. In particular, biophysical modeling provides a principled approach to deciphering how gene regulatory programs are encoded in DNA and RNA sequences \cite{Kinney:2010tn, Segal:2009jv, Belliveau:2018kr, Ireland:2020aa, Ishigami:2024aa}. Our work provides a formal language in which such biophysical models can be expressed and then analyzed. This capability may allow researchers to systematically explore the space of biophysical models of gene regulation, thereby automating the construction and inference of biophysical models for gene regulatory codes.

\begin{acknowledgments}
We thank Rob Phillips for his support and encouragement throughout this project, as well as Muir Morrison for his early work with JBK on this topic. RJR further thanks Sergei Gukov and Alexei Kitaev for helpful discussions. The work of JBK was supported by NIH grants GM133777 and HG011787. The work of RJR was supported by NIH grant GM118043. This research was performed in part at Aspen Center for Physics, which is supported by NSF grant PHY-2210452.
\end{acknowledgments}

\appendix
\section{Algebra of mode and field operators}\label{app:operatoralgebra}

Since each mode represents a hard-core boson, mode-specific creation and annihilation operators are nilpotent, i.e., 
\begin{align}
    \hat{A}_{i}^2 = \check{A}_{i}^2 = 0.
\end{align}
When multiplied by presence and absence operators for the same mode, the creation and annihilation operators are readily seen to satisfy
\begin{align}
\bar{A}_i \hat{A}_i& = \hat{A}_i \tilde{A}_i = \hat{A}_i,~~~~\tilde{A}_i \hat{A}_i = \hat{A}_i \bar{A}_i = 0, \\
\tilde{A}_i \check{A}_i &= \check{A}_i \bar{A}_i = \check{A}_i,~~~~\bar{A}_i \check{A}_i = \check{A}_i \tilde{A}_i = 0.
\end{align}
Presence and absence operators are both idempotent, i.e.,
\begin{equation}
\bar{A}_i^2 = \bar{A}_i,~~~\tilde{A}_i^2 = \tilde{A}_i,
\end{equation}
and mixed products of presence and absence operators for the same mode vanish:
\begin{equation}
\bar{A}_i \tilde{A}_i = \tilde{A}_i \bar{A}_i = 0.
\end{equation}
From these properties and the fact that operators for distinct modes commute, we get the following commutation relations for mode operators:
\begin{align}
    [\check{A}_i, \hat{A}_j] &= \delta_{ij}(\tilde{A}_i - \bar{A}_i) = \delta_{ij}(1 - 2 \bar{A}_i), \\
    [\bar{A}_i, \hat{A}_j] &= [\hat{A}_i, \tilde{A}_j] = \delta_{ij}\hat{A}_i, \\
    [\tilde{A}_i, \check{A}_j] &= [\check{A}_i, \bar{A}_j] = \delta_{ij}\check{A}_i. 
\end{align}
Summing over indices gives the corresponding commutation relations for field operators:
\begin{align}
    [\check{A}, \hat{A}] &= \tilde{A} - \bar{A} = N - 2 \bar{A}, \\
    [\bar{A}, \hat{A}] &= [\hat{A}, \tilde{A}] = \hat{A}, \\
    [\tilde{A}, \check{A}] &= [\check{A}, \bar{A}] = \check{A}. 
\end{align}
These commutation relations, together with $\bar{A}\ket{0} = 0$, allow us to compute the impact of each field operator on the macrostate $\ket{n}$,
\begin{align}
    \hat{A}\ket{n} 
        &= \hat{A}\frac{\hat{A}^n}{n!}\ket{0} \nonumber\\
        &= (n+1)\frac{\hat{A}^{n+1}}{(n+1)!}\ket{0}\nonumber\\
        &= (n+1)\ket{n+1},\label{eq:AppAophatnstate} \\
    \bar{A}\ket{n} 
        &= \bar{A}\frac{\hat{A}^n}{n!}\ket{0} \nonumber\\
        &= \frac{1}{n!} [\bar{A}, \hat{A}^n] \ket{0} \nonumber\\
        &=\frac{1}{n!}\sum_{k=0}^{n-1}\hat{A}^{n-k-1}[\bar{A},\hat{A}]\hat{A}^k\ket{0}\nonumber\\
        &= n\frac{\hat{A}^n}{n!}\ket{0} = n\ket{n},\label{eq:AppAopbarnstate} \\
    \check{A}\ket{n} 
        &= \check{A}\frac{\hat{A}^n}{n!}\ket{0} = [\check{A},\frac{\hat{A}^n}{n!}]\ket{0} \nonumber\\
        &= \frac{1}{n!}\sum_{k=0}^{n-1}\hat{A}^{n-k-1}[\check{A},\hat{A}]\hat{A}^{k}\ket{0} \nonumber\\
        &= \frac{1}{n!}\sum_{k=0}^{n-1}\hat{A}^{n-k-1}(N-2\bar{A})\hat{A}^{k}\ket{0} \nonumber\\
        &= \frac{1}{n!}\sum_{k=0}^{n-1}(N-2k)\hat{A}^{n-1}\ket{0} \nonumber\\
        &=\frac{1}{n!} \left(nN - 2 \frac{n(n-1)}{2} \right) \hat{A}^{n-1} \ket{0} \nonumber\\
        &= (N-n+1)\frac{\hat{A}^{n-1}}{(n-1)!}\ket{0}\nonumber\\
        &= (N-n+1)\ket{n-1},\label{eq:AppAopchecknstate} \\
    \tilde{A}\ket{n} 
        &= \tilde{A}\frac{\hat{A}^n}{n!}\ket{0} \nonumber\\
        &= (N-\bar{A})\frac{\hat{A}^n}{n!}\ket{0} \nonumber\\
        &= (N-n)\frac{\hat{A}^n}{n!}\ket{0} \nonumber\\
        &= (N-n)\ket{n}.\label{eq:AppAoptildenstate}
\end{align}

\section{Non-commutation of monomer and dimer operators}\label{app:noncommutation}

Here we derive the expression for $[\hat{M}, \check{D}]$ in \eq{eq:Mhat_Dcheck_commutator}. We begin by evaluating the commutator on the individual composite mode operators:
\begin{align}
    [\hat{M}_k,\check{D}_{ij}] 
    &= \check{I}_{ij} \left[ \hat{A}_k \tilde{a}_k, \check{A}_i \check{A}_j \check{a}_i \check{a}_j \right] \nonumber \\
    &= \delta_{ki} \check{I}_{ij} \check{A}_j \check{a}_j \left[ \hat{A}_k \tilde{a}_k, \check{A}_i  \check{a}_i \right] \nonumber \\
    &~~~~+ \delta_{kj} \check{I}_{ij} \check{A}_i \check{a}_i  \left[ \hat{A}_k \tilde{a}_k, \check{A}_j  \check{a}_j \right]. \label{eq:appendixB_eq1}
\end{align}
Considering the commutator in the $k=i$ term and dropping the subscripts for brevity, we find that
\begin{align}
    [\hat{A} \tilde{a}, \check{A} \check{a}] &= \hat{A} [\tilde{a},\check{A}] \check{a} + [\hat{A},\check{A}]\tilde{a} \check{a} + \check{A} \hat{A} [\tilde{a},\check{a}] + \check{A} [\hat{A},\tilde{a}] \check{a} \nonumber \\
    &= \bar{A} \check{a} \nonumber \\
    &= \hat{A} \tilde{a} \check{A} \check{a}.
\end{align}
The same holds for the $k=j$ term. Substituting these back into \eq{eq:appendixB_eq1} and summing over $k$ gives 
\begin{align}
    \sum_k [\hat{M}_k,\check{D}_{ij}] 
    &=  \sum_k \delta_{ki} \hat{A}_i \tilde{a}_i \check{I}_{ij} \check{A}_i \check{a}_i \check{A}_j \check{a}_j  \nonumber \\
    & +  \sum_k \delta_{kj} \hat{A}_j \tilde{a}_j \check{I}_{ij} \check{A}_i \check{a}_i \check{A}_j \check{a}_j \nonumber  \\
    &= (\hat{M}_i + \hat{M}_j) \check{D}_{ij}.
\end{align}
From this we recover \eq{eq:Mhat_Dcheck_commutator}.

\section{Factory/gallery equivalence for the homopolymer}\label{app:FGequivpol}

We now prove the equivalence of the factory and gallery representations for the homopolymer system in Section \ref{sec:homopol}. As with the homodimer, we do this by evaluating individual terms in the Taylor expansion of $e^{\mathbb{F}_2} e^{\mathbb{F}_1} \ket{0}$, with factory operators defined as in \fig{fig:homopolymer_specification}{(a)}. By inspection we see that all full contractions of $\mathbb{F}_2^p \mathbb{F}_1^q$ must consist of a product of $x$-chain ($\hat{C}_x$) and $x$-ring ($\hat{R}_x$) operators. Wick's theorem therefore gives
\begin{equation}
	\mathbb{F}_2^p\, \mathbb{F}_1^q \ket{0} = \!\!\!\!\sum_{\set{c_x, r_x} | p,q} \!\! \Omega^{(p,q)}_{\set{c_x,r_x}} \prod_{x=1}^\infty \hat{C}_x^{c_x} \hat{R}_x^{r_x} \ket{0},
\end{equation}
where $c_x$ denotes the number of $x$-chains, $r_x$ denotes the number of $x$-rings, $\set{c_x, r_x} | p,q$ denotes all sets of these numbers that are consistent with $p$ bonds and $q$ particles, i.e., which satisfy 
\begin{equation}
	q = \sum_{x=1}^\infty (x c_x + x r_x), ~~~~p = \sum_{x=1}^\infty ([x-1] c_x + x r_x),
\end{equation}
and $\Omega^{(p,q)}_{\set{c_x,r_x}}$ is a combinatorial coefficient that quantifies the number of distinct contractions that yield $\set{c_x, r_x} | p,q$. 

We now compute $\Omega^{(p,q)}_{\set{c_x,r_x}}$. The number of ways to partition $q$ monomers among $c_x$ distinct $x$-chains and $r_x$ distinct $x$-rings is given by
\begin{equation}
\omega_q = \frac{q!}{\prod_{x=1}^\infty (x!)^{c_x + r_x}}.
\end{equation}
Similarly, the number of ways to partition $p$ bonds among the $x$-chains and $x$-rings is
\begin{equation}
\theta_p = \frac{q!}{\prod_{x=1}^\infty ([x-1]!)^{c_x}(x!)^{r_x}}.
\end{equation}
Since one can rearrange the $x$-chains among themselves and the $x$-rings among themselves without changing the result, the number of unique partitions is the product of $\omega_q$ and $\theta_q$ divided by an exchange factor of $\pi_{p,q} = \prod_{x=1}^\infty c_x! r_x!$. Moreover, there are $\sigma_x = (x-1)! x!$ distinct ways of constructing each $x$-chain from a given set of $x$ particles and $x-1$ bonds, and $\rho_x = x! x!$ ways to construct each $x$-ring from a set of $x$ particles and $x$ bonds. Note that the circular symmetry, which contributes a factor of $1/x$ to this second quantity, is already accounted for in the definition of the $x$-ring and should not be double-counted here. We therefore find that
\begin{align}
	\Omega^{(p,q)}_{\set{c_x,r_x}} = \frac{\omega_q \theta_p}{\pi_{p,q}} \prod_{x=1}^\infty \sigma_x^{c_x} \rho_x^{r_x} = \frac{p! q!}{\prod_{x=1}^\infty c_x! r_x!}.
\end{align} 
Consequently, 
\begin{align}
	e^{\mathbb{F}_2} e^{\mathbb{F}_2} \ket{0} & = \sum_{p,q} \frac{1}{p! q!} \sum_{\substack{\set{c_x, r_x} | p,q}} \frac{p! q!}{\prod_x c_x! r_x!}  \prod_x \hat{C}_x^{c_x} \hat{R}_x^{r_x} \ket{0} \nonumber \\
	& = \sum_{c_1, c_2, \ldots} \sum_{r_1, r_2, \ldots} \prod_x \frac{\hat{C}_x^{c_x}}{c_x!} \frac{\hat{R}_x^{r_x}}{r_x!} \ket{0} \nonumber \\
    &= e^{\sum_x (\hat{C}_x + \hat{R}_x)} \ket{0}. 
\end{align}
This establishes the factory/gallery equivalence for the homopolymer.

\section{Derivation for the species-specific depletion operator}\label{app:depderiv}
Here we derive the species-specific depletion operator in \eq{eq:Qgrav}. First we express the species-specific reaction operator as
\begin{align}
    \mathbb{Q} = \prod_{k=1}^K P_k Q_k,~~\textrm{where}~~
    P_k = \frac{\hat{G}_k^{p_k}}{p_k!},~~Q_k = \frac{\check{G}_k^{q_k}}{q_k!}.
\end{align}
The assumption of non-overlapping $\vec{p}$ and $\vec{q}$ implies that $P_k = 1$ and/or $\mathbb{Q}_k = 1$ for all $k$. Because of this, 
\begin{align}
    \grave{\mathbb{Q}} = \prod_{k=1}^K \grave{P}_k \grave{Q}_k.
\end{align}
Note that this will generally not be true if any $P_k$ and $Q_k$ are both non-unity, since the depletion version of a product of operators for a given field is generally not the product of the depletion version of each operator. 

Next we express $P_k$ and $Q_k$ in terms of mode-specific operators. Using the fact that the mode-specific creation and annihilation operators are nilpotent (and dropping $k$ to ease notation), 
\begin{align}
    P = \sum_{\mathcal{I}:|\mathcal{I}|=p} \prod_{i \in \mathcal{I}} \hat{G}_i,~~~
    Q = \sum_{\mathcal{I}:|\mathcal{I}|=q} \prod_{i \in \mathcal{I}} \check{G}_i.
\end{align}
Transforming $\hat{G}_i \to \tilde{G}_i$ and $\check{G}_i \to \bar{G}_i$, we obtain the corresponding depletion operators 
\begin{align}
    \grave{P} = \sum_{\mathcal{I}:|\mathcal{I}|=p} \prod_{i \in \mathcal{I}} \tilde{G}_i,~~~
    \grave{Q} = \sum_{\mathcal{I}:|\mathcal{I}|=q} \prod_{i \in \mathcal{I}} \bar{G}_i.
\end{align}
Applying each of these operators to a macrostate $\ket{n}$, one finds that
\begin{align}
    \grave{P}\ket{n} &= {N-n \choose p}\ket{n} = {\tilde{G} \choose p} \ket{n}, \\
    \grave{Q}\ket{n} &= {n \choose q}\ket{n} = {\bar{G} \choose q} \ket{n},
\end{align}
and therefore,
\begin{align}
    \grave{P} = {\tilde{G} \choose p},~~~~\grave{Q}={\bar{G} \choose q}.
\end{align}
Reintroducing the $k$ subscripts, we obtain \eq{eq:Qgrav}:
\begin{align}
     \grave{\mathbb{Q}} = \prod_{k=1}^K \grave{P}_k \grave{Q}_k = \prod_{k=1}^K {\tilde{G}_k \choose p_k}{\bar{G}_k \choose q_k}. \label{eq:Q_grave}
\end{align}

\section{Code availability}\label{app:code}
Python code implementing Algorithm \ref{alg:one} as well as the Jupyter Notebooks used to perform the simulations in \fig{fig:simulations}{} and \fig{fig:various_simulations}{} are available at \href{https://github.com/Rebecca-J-Rousseau/RousseauKinney2024_algebra}{github.com/Rebecca-J-Rousseau/RousseauKinney2024\_algebra}.

\bibliography{main}

\end{document}